\documentstyle[emulateapj,psfig,epsf,epsfig,apjfonts]{article}

\newcommand{\mum}{$\,\mu$m}

  \def\itm#1 {\vskip10pt \noindent \square\ {\bf #1} }
  \def\square {\hbox{\vrule width5pt height5pt}}

\def\gs{\mathrel{\raise0.35ex\hbox{$\scriptstyle >$}\kern-0.6em 
\lower0.40ex\hbox{{$\scriptstyle \sim$}}}}
\def\ls{\mathrel{\raise0.35ex\hbox{$\scriptstyle <$}\kern-0.6em 
\lower0.40ex\hbox{{$\scriptstyle \sim$}}}}
\lefthead{Chapman et al.}
\righthead{The optically faint radio sources}

\begin{document}

\title{Understanding the nature of the optically faint radio sources 
and their connection to the submillimeter population}

\author{Scott C.\ Chapman,$\!$\altaffilmark{1} 
Geraint F.\ Lewis,$\!$\altaffilmark{2} 
Douglas Scott,$\!$\altaffilmark{3} Colin Borys,$\!$\altaffilmark{3} 
Eric Richards$\!$\altaffilmark{4}
}
\affil{California Institute of Technology,
Pasadena, CA 91125,~~U.S.A.}
\affil{Anglo-Australian Observatory, P.O. Box 296, Epping, 
	NSW 1710, Australia}
\affil{Department of Physics \& Astronomy, 
University of British Columbia, V6T~1Z1, Canada}
\affil{Department of Physics \& Astronomy, 
University of Alabama, Huntsville, 85287, U.S.A.}



\begin{abstract}
We present a sample of 43 submillimeter sources detected (at ${>}\,3\sigma$),
drawn from our program to follow-up optically faint radio sources with SCUBA.
These sources already have associated radio 
and in many cases optical identifications, and many are
also detected at 450\mum.
We compare these with 12 submillimeter sources drawn from the literature,
which were discovered in blank field mapping campaigns, but also
have radio detections.  We then use this total sample (55 sources) 
to study and model the evolution of dusty galaxies.
A correlation is observed in the sub-mm/radio 
color-magnitude diagram, which can be modeled by strong luminosity evolution.
The selection effects of the radio/optical 
pre-selection technique are determined from the models, and a corrected
redshift distribution is constrained using a range of model assumptions.
The temperature/redshift effects on the 450\mum\ detected subset of our
sample are studied in relation to the models, and
prospects for improved measurements in the shorter sub-mm
wavelength windows (450\mum\ and 350\mum) are explored.

\end{abstract}

\keywords{galaxies: evolution --- galaxies: formation --- sub-mm: galaxies
--- radio: galaxies}

\section{Introduction}

The advent of the Sub-millimeter Common User Bolometer Array (SCUBA) on the
James Clerk Maxwell Telescope (JCMT) has spawned a new era of discovery 
in the sub-mm wavebands. Sub-mm luminous, extragalactic 
sources were quickly uncovered (Smail, Ivison \& Blain 1997)
as the possible high redshift analogs to 
local Ultra-Luminous InfraRed Galaxies (ULIRGs).  However, the population 
continues to be poorly understood, largely as a result of two 
observational difficulties. Firstly,
obtaining large samples of objects from SCUBA sub-mm mapping is an arduous 
process, whereby typically one to two sources are uncovered in a night's
worth of integration on a blank piece of sky.
Secondly, identifying secure detections at other wavelengths is
tremendously difficult due to the positional uncertainty and 
large SCUBA beam-size (15 arcsec),
and the intrinsic faintness of most sources at all other wavelengths.

The tight correlation observed locally between thermal far-IR emission and
synchrotron radio emission (Helou et al.~1985, Condon 1992) suggests a
possibility for identifying the sub-mm sources. The positional accuracy
and small beamsizes of large radio interferometers like the VLA can act as
a surrogate to the sub-mm, allowing precise identifications at other
wavelengths. Smail et al.~(2000) demonstrated that a significant fraction 
of their sub-mm sources could be detected in the radio.
However, reaching the radio depths required to detect the
sub-mm sources is not trivial.  Even the deepest radio map yet obtained, a
100 hour VLA integration at 1.4\,GHz in the SSA13 survey region
(Fomalont et al.~2002), cannot
detect all the bright sub-mm sources found in this field 
(Barger et al.~2001a).  At this depth ($S_{1.4 GHz}$=20\,$\mu$Jy)
sub-mm sources with warm (${\sim}\,50\,$K) dust and
luminosities similar to the local ULIRG, Arp220 
will likely be missed at $z\,{\ga}\,3$.
Colder sources would be missed at even lower redshifts (Blain 1999).

Putting aside the precise form of the radio/sub-mm overlap,
deep radio observations have emerged as an efficient means to 
pre-select high redshift, sub-mm sources 
(Barger, Cowie, Richards 2000, hereafter BCR; Chapman et al.~2001a,
hereafter C01).  This
effectively circumvents the time
constraint for obtaining large samples of objects with the present
generation sub-mm imaging instruments, SCUBA (Holland et al.~1999)
and MAMBO (Bertoldi et al.~2000).
These surveys (BCR, C01) 
have used the bi-modal break in the apparent optical magnitudes of
radio source identifications lying at $I\sim24$ (Richards et al.~1999),
selecting only the relatively faint sources for sub-mm follow-up
-- the Optically Faint Radio Sources (OFRS).
While the entire blank field sub-mm population (e.g.~Smail et al.~2001)
is not selected in this manner, a significant percentage of the
source counts are recovered (${\sim}\,$70\%, BCR, C01).
The key advantage, however,
is that the optical and radio properties of the sources are known with
certainty, with the high resolution radio providing
morphological information for studying the types of galaxies
in the sub-mm population (Richards~2000).
The challenge is to understand the selection function sufficiently to
exploit the rich information gleaned by the large samples of sub-mm
sources uncovered in this manner.

In this paper, we investigate 
the evolutionary behavior of the sub-mm population of galaxies,
constrained by the correlations observed in a large sample selected using the
radio.  We develop models to describe the form of the evolution and 
study the model-implied radio selection function.
The detailed, multi-wavlength 
properties of this large sample of radio selected sub-mm
sources are presented elsewhere (Chapman et al.~in preparation, Barger et 
al.~2001b).   We give only a brief description in Section~2, and show
color-magnitude diagrams in Section~3.
We adopt two modeling approaches.   In Section~4 we take the local {\sl IRAS\/}
bright galaxy sample and evolve individual sources according to prescriptions
which are able to fit the SCUBA counts (e.g.~Blain et al.~1999a) and the
far-infrared background (Puget et al.~1996, Fixsen et al.~1998).
In Section~5 we adopt a Monte Carlo approach, drawing luminous infrared 
galaxies randomly from evolving distributions.
We use these models to better understand the range
in properties sampled by the radio pre-selected sub-mm galaxy population
in Section~6.  We then discuss shorter wavelength sub-mm detections in
Section~7 and consider wider implications and future directions in Section~8.

\section{Sample definition}

We draw on the large number of sub-mm sources cataloged and described
in Chapman et al.~(in preparation) and Barger et al.~(2001b),
which were obtained through SCUBA follow-up to 
radio sources in the extended Hubble Deep Field (HDF), and the
Hawaii SSA13 survey field.  The main sample consists of 
43 sources detected at 850\mum\ above 3 times the RMS, corresponding to
about 5\,mJy.  Twenty of these were already presented in BCR and C01.
The sources were observed in SCUBA `photometry' mode, keeping the
source on a bolometer at all times (3-bolometer chopping -- C01) in order
to go deeper than making a fully sampled map of the field.
Targets were selected to be optically faint, with no optical 
source  having flux $I<24$ being coincident with the radio position.

Several of these sources have radio/optical astrometry which is difficult 
to reconcile, with radio source centroids offset from relatively bright optical 
sources by 0.5\arcsec\ to 1.5\arcsec. In 
C01, we assumed that 6 such sources were optically fainter than $I>25$
(3 detected in the sub-mm, and 3 undetected).  Further statistical analysis
and a more precise astrometric solution now suggest that these optically 
brighter sources {\it are\/} the likely counterparts to the radio sources.
Sub-mm follow-up of radio sources subsequent to the C01 study
has extended the sample into optically brighter sources as described
in Chapman et al.~(2002a). Such endeavors have continued to uncover
sub-mm luminous sources, albeit at a lower recovery rate than the $I>25$
samples.   We estimate that the sub-mm recovery rate for radio sources with
$I>24$ is 38\%.

The sub-mm follow-up to radio sources leaves a large {\it undetected\/}
sample, with an implied $S_{850}<5\,$mJy. 
This begs the question as to the nature of these sources.
As discussed in C01, this undetected radio sample has 
significant sub-mm flux on average, even if individually they are too
faint to detect.  Since there is no significant 
difference in the radio flux distribution between the 
sub-mm detected and undetected OFRS samples, one interpretation is 
that they are a lower redshift continuation of our detected sample
(intrinsically less radio-luminous and harder to detect with SCUBA).
However, we do not discount the possibility raised in Richards et al.~(1999)
that these sources could be high redshift radio-loud AGN.  Some fraction
of them could also be displaced radio lobes from other radio sources.

A simple inspection of the SED for the ULIRG Arp220 suggests that the
deep VLA maps used both to select the OFRS for SCUBA follow-up and to 
attempt identification of sub-mm blank field surveys (${\sim}\,40\,\mu$Jy
completeness) will not detect a galaxy with twice the luminosity
of Arp220 beyond $z\,{\sim}\,3$ ($S_{850 \mu m}\sim5\,$mJy). 
However, the sub-mm selection function suggests that
sources typically detectable by SCUBA should lie mostly at
$z\,{>}\,1$ (Blain et al.~1999b). Thus the radio identified sub-mm
population suggests a redshift range spanning $z\,{\sim}\,$1--3.

We also collected sub-mm sources with radio identifications from the
literature, 
in order to compare the properties of sources drawn from blank fields to 
our OFRS selected sub-mm sources. Seven sources from the Smail et al.~(2001)
catalog and five from Eales et al.~(2000) were identified in the radio
and can thus be compared with our sample.
Smail et al.~(2001) have suggested optical counterparts for their
objects, 4 of which are optically fainter than $I\,{>}\,24$,
while 3 lie between $21\,{<}\,I\,{<}\,24$.

%
%
%

\section{The sub-mm/radio color-magnitude diagram}

A favorable situation exists for studying high redshift sources in the
sub-mm waveband:
for sub-mm sources lying at redshifts greater than about one,
we can use the sub-mm flux alone to estimate the overall
luminosity, only weakly dependent on redshift because
of the offsetting effects of the steep positive spectral index in
the sub-mm and cosmological dimming.
The negative spectral index in the radio further implies that the
sub-mm/radio ratio can be used as a rough
redshift indicator (Carilli \& Yun, 1999,2000; Dunne et al.~2000; BCR),
effectively tracing the trough in the spectral
energy distribution (SED) formed by the opposing spectral indices. 

The flat luminosity-redshift relation implies that any trends observed
in the ratio of sub-mm/radio as a function of sub-mm flux
must be a result of evolutionary effects. 
Of immediate interest for our sample is therefore to
analyze the $S_{850\mu{\rm m}}$/$S_{1.4{\rm GHz}}$ 
versus $S_{850\mu{\rm m}}$ plane -- essentially a sub-mm/radio 
color-magnitude diagram (CMD).  For sources at $z\,{>}\,1$ and in 
the absence of varying dust effects,
such a CMD can be interpreted simply in terms of redshift ($y$-axis) and
luminosity ($x$-axis).  However, the redshift estimated in this manner
is degenerate with dust temperature (Blain 2000) and it is only the quantity
$(1+z)/T_{\rm d}$ which is truly constrained by the sub-mm/radio ratio.
This increases the impetus to study the CMD 
in order to assess any obvious trends of this redshift/temperature indicator
with sub-mm flux density.

In Fig.~1, we present the sub-mm/radio CMD for the total sub-mm observed
OFRS sample.
Red circles represent radio sources detected with ${\ga}\,3\sigma$
significance
at 850\mum. An error bar representative of the RMS error for the detected
sample is shown in the lower right corner.
Blue circles show an additional 24 sources from our sample 
which have ${\ga}\,2\sigma$ significance at 850\mum.
While we do not consider these robust detections,
a large fraction will be real sub-mm sources, and
the average properties of this {\it marginally\/} detected sample will
help extend the statistical properties of OFRSs to fainter fluxes.
Crosses represent radio sources with $<2\sigma$ significance at 850\mum,
where sources with negative sub-mm flux density
have been arbitrarily plotted at $S_{850}=0.3$\,mJy.

In interpreting Fig.~1 we caution that our sample is only complete 
at a flux limit of $S_{850}\,{\ga}\,5\,$mJy, resulting in accurate
distributions in the CMD plane only for these brightest sources. 
Nevertheless
an apparent correlation is seen, whereby an increasing ratio of
850\mum\ flux to 1.4\,GHz flux scales with increasing $S_{850}$.  This
correlation plausibly extends into 
the regime of radio sources with $<3\sigma$ significance in the sub-mm
(open circles and crosses).  However,
the absence of sources in the upper left corner of the diagram is largely
the result of the radio flux limit of the survey, which we represent
with a dashed line.  Note that the two ${>}\,3\sigma$
sources lying above the line were selected from the deeper 
SSA13 radio field.  On the other hand, the dearth of sources in the lower
right of Fig.~1, compared with the high density of sources in the
upper right, must be a true evolutionary effect.

As a comparison with other studies, we also plot in Fig.~1 the sub-mm
sources with radio counterparts from the catalogs of Smail et al.~(2001)
and Eales et al.~(2000) as 7 triangles and 5 squares respectively.
Of interest is the fact that these sources lie scattered around
the complete CMD area probed by our OFRS sample, 
with no obvious bias for the optically brighter members.
It is therefore plausible that these sources may trace the same evolutionary
trends uncovered by our OFRS population, irrespective of the optical emission.
We return to this point later.

Note that all sources presented in the diagram are robustly detected in the
radio, and are likely to be relatively high star formation rate galaxies as
noted in section 2 above, 
whether or not we were able to obtain secure sub-mm detections. 
We can thus study with confidence the CMD at the bright end.
However, a more detailed assessment of the population
lying in the 1--$5\,$mJy regime will require additional, deeper sub-mm 
measurements, which would provide
significant leverage of the CMD correlation to fainter sub-mm flux densities.

\section{Local CMD relations and a simple picture of evolution}

To better understand the intrinsic scatter and correlations present in the
sub-mm/radio population at higher redshifts, we examine the local
infrared luminous population in the radio and at sub-mm to far-IR wavelengths.
We present a sample of 19 objects from the {\sl IRAS\/}
bright galaxy sample (BGS) which have been
measured at 850 and 450\mum\ (Dunne \& Eales 2001), and in the radio (Condon 
et al.~1996).  We plot them in the CMD plane (Fig.~2a), 
as well as a color-color diagrams (CCDs) consisting of
either $S_{100}/S_{850}$ or $S_{450}/S_{850}$ versus $S_{850}/S_{1.4{\rm GHz}}$
(Fig.~2b,c).
For the range of likely redshifts of the SCUBA sources in this study
($z\sim$1--4), the 850\mum\ band will remain a probe of the Raleigh-Jeans
grey-body tail of the SED.  However, the observed 450\mum\ band will sample
a very different part of the SED at higher redshifts, comparable to
${\sim}\,100$\mum\ locally.

The 850\mum/radio CMD reflects the intrinsic scatter in the far-IR/radio
relation, coupled with the dispersion from dust temperatures and
emissivities, as well as measurement uncertainties.  The measured 
scatter of 0.2 dex in 850\mum/radio (Table~1) is close to that found for the
far-IR/radio (Helou et al.~1985).

The 450\mum/850\mum\ CMD shows a flat relation, representing a fairly constant
and cool dust temperature component (${\sim}\,20$\,K), as discussed in 
Dunne \& Eales (2001).
The 100\mum/850\mum\ and 60\mum/100\mum\ CMDs, however, 
show a rather large scatter, indicative of a range in 
warmer ($\sim$40\,K) dust properties found in this local infrared luminous
sample.

The local galaxy 
color-color relations, using the 450\mum/850\mum\ versus 850\mum/radio, 
continue to reflect the
apparent constancy of the cool dust properties of this sample.
The color-color 
relation using the local 100\mum\ band, however, now forms a strong
correlation (compare Figs.~2a,b), 
reflecting the effect of dust temperature and emissivity variations
on both the 100\mum/850\mum\ and the 850\mum/radio.
This arises since the change in temperature effectively changes the total
far-IR luminosity for a fixed 850\mum\ measurement, and thus changes the
correlation of far-IR/radio.
The 60\mum\ band is similar to the 100\mum, but exhibits larger scatter as it
lies beyond the grey-body peak and hence has contributions from a range of
dust components.

While 19 objects is a small sample from which to establish local correlations,
a larger sample exists for {\sl IRAS\/} galaxies measured only at 850\mum\
Dunne et al.~(2000a).  Color-color diagrams for this larger {\sl IRAS\/}
sample are presented in Fig.~2c. 
The steep negative correlations in the 100\mum/850\mum\ and 60\mum/850\mum\
diagrams are still observed, with similar scatter as in the 19 object sample. 
For reference, we list the scatter and slope of all these local correlations
in Table~1

These correlations may arise due to
sources with the highest dust temperatures (and hence 100\mum/850\mum,
60\mum/100\mum\ ratios) having
effectively have lower FIR fluxes, for galaxies in this limited range
of $L_{850}$. The 850\mum/$1.4\,$GHz ratio, scaling inversely with far-IR,
is therefore also lower.
The 850\mum/$1.4\,$GHz correlation from the SED template catalog
of Dale et al.~(2001a,b) suggests that
the more quiescent galaxies have
a significantly greater 850\mum/$1.4\,$GHz ratio than active IR-luminous
star forming galaxies (as expected since the quiescent galaxies have
colder dust).  By redshifting the SEDs, we can anticipate the effect
of this monotonic scaling of 850\mum/$1.4\,$GHz with galaxy IR luminosity
on our high-$z$ CMD and CCD relations.
Local infrared sources can thus be understood as correlating tightly in
color-color space, due to the effect of dust properties and star-formation
on both the radio and the far-IR windows, leaving the 850\mum\ and 450\mum\
flux densities relatively constant.
This behavior is modeled in detail in Dale et al.~(2001a,b).
 
In the absence of evolution, tracing sources to higher redshifts would lead to
results following the trend discussed above: the sources retain the
same observed flux density, whilst increasing in 850\mum/radio as a function
of redshift (Fig.~3a).
If we propose instead that a population of IR-luminous objects undergoes
a strong luminosity evolution of the form $(1+z)^4$ out to $z\,{=}\,3$,
turning over to $(1+z)^{-4}$ to $z\,{=}\,6$ (Fig.~3b), then
the local sources move along the direction suggested
in our observed SCUBA source sample (Fig.~1). 
The CMD trend can be understood naturally in terms of luminosity evolution:
more luminous objects occur at higher redshifts, leaving the
lower right corner largely empty.  Note that here, and in what follows, we
explicitly use an Einstein-de Sitter cosmology.

The source density in comoving coordinates is fixed, since
we have simply evolved the local sample of sources in luminosity.
The $z\,{=}\,0$ sources (shown by the point symbols in Fig.~3) would therefore
only be detected in an enormous survey field.
A dashed line represents the radio sensitivity limit of the VLA for
${\sim}\,50$~hour integrations at $1.4\,$GHz.  Sources beyond the peak in our 
luminosity evolution function quickly traverse this radio limit and would
not be detected either.
The luminosity range of these sources spans $\log(L_{\rm FIR})$ = 10 to 12. 
For the
luminosity evolution invoked in our model, all the sub-mm sources
could evolve from such a local sample and be at redshifts 1--4.

The local source sample represents an average dust temperature of
$T_{\rm d}\,{=}\,36$\,K (Dunne et al.~2000a).
In Fig.~3c, we also evolve a subset of the local sources with cooler 
($T_{\rm d}\,{=}\,25$\,K -- triangles) and hotter ($T_{\rm d}\,{=}\,50$\,K
-- open circles) dust temperatures, keeping the 850\mum\
luminosity fixed.  As demonstrated in Dale et al.~(2001a,b), this is
equivalent to fixing the mid-IR luminosity (${\sim}\,10\,\mu$m), but results in 
proportionately higher far-IR (hot dust) or lower far-IR (cool dust) sources.
Therefore a sample of equivalently far-IR luminous sources with colder dust
temperatures will be detected by SCUBA at much lower redshifts, and a
less extreme luminosity evolution would be required to generate sources
which lie in our sub-mm detection window $S_{850}\,{>}\,5\,$mJy.
The effect of moving to a flat $\Lambda=0.7$ cosmology, however, is to 
introduce a slope to the CMD which also has to be countered with
luminosity evolution or colder sources to maintain sources in our detection
window.  However, the evolutionary effects are sufficiently strong that the
precise cosmology adopted is relatively unimportant.

\section{Modeling the evolving CMD relations}

We now embark on more comprehensive models of our sub-mm/radio CMD and the 
trends observed in the previous sections. 
Our aim will be to simulate the properties of our data sample using
a Monte Carlo approach, and to use this to focus on the evolutionary trends
implied by the SCUBA data for the OFRS sample.  The data we have do not
justify embarking upon a detailed fitting of various model parameters.  Rather
we look at some examples of models which are motivated by fitting what is
already known about local ULIRGs and distant SCUBA sources, and using these
examples try to draw general conclusions.
We start with simple step by step assumptions about the 
population and progress to more involved and (hopefully) realistic
effects, taking careful note of the changes in the resulting CMD.

Monte Carlo simulations of the 850\mum/1.4\,GHz CMD are first attempted 
by assuming a population of Arp220-like SEDs, equivalent in the
far-IR/sub-mm regime to a greybody with dust temperature
$T_{\rm d}\,{=}\,50\,$K and emissivity $\beta\,{=}\,1.5$.
Fig.~4a depicts an ensemble of such objects in the CMD, all having the
same luminosity (approximately twice that of Arp220).
The sources follow
a Gaussian redshift distribution (indicated in the insets in Fig.~4)
matched to the spectroscopically estimated
distribution of Barger et al.~(1999), thought to be plausible in light
of recent Keck redshift measurements for sub-mm sources (Chapman et al.~2002b).
The sources are color-coded by redshift as described in the
figure caption, and represent density evolution of a single 
luminosity of source.  This is effectively the opposite hypothesis to our
toy model above, where the source density per comoving volume remained
constant for all time, but evolved in luminosity. 
The corresponding track then
traces this characteristic sub-mm galaxy as a function of redshift
through the CMD.  Note that beyond $z\,{\sim}\,1$, the source is observed
with approximately the same 850\mum\ flux density for the adopted
$\Omega_{\rm M}\,{=}\,1$
cosmology, reflecting the favorable sub-mm $K$-correction described above.
However, redshifts map monotonically onto the 850\mum/$1.4\,$GHz axis.

Since we know the local scatter in the 850\mum/$1.4\,$GHz relation (as
discussed in Section~4, see also Helou et al.~1995, Yun et al.~2001), 
we can introduce the assumption that this scatter remains constant
and applies at all redshifts.  We describe the relation as a Gaussian in
the 850\mum/$1.4\,$GHz ratio with an RMS scatter of 0.2 dex. 
The effect of the scatter of this relation on the CMD distribution is
shown in Fig.~5b, distributing the single luminosity sub-mm family vertically.
Without spectroscopic measurements, our current data cannot constrain the
true evolution in the far-IR/radio correlation.  Such 
information will have to await future
redshift surveys of radio and sub-mm detected
sources (e.g.~Chapman et al.~2002b).  At the moment assuming it is the same
as the local relation is about the best that can be done.

We can alternatively impose a luminosity distribution on our
ULIRG population, as represented in Fig.~4c.
The bright SCUBA counts at 850\mum\ are well fit by a power law
(e.g.~Barger et al.~1999), and hence we first adopt a constant power law
for our luminosity functions over all redshifts (shown in the lower insets
in Fig.~4).  This is probably
a reasonable assumption for the bright sub-mm sources considered here.
This therefore represents a null hypothesis for luminosity evolution.
At this point, our radio survey completeness of $40\,\mu$Jy imposes a diagonal
cut on the sources from the upper left corner of the CMD.

We can then combine both the farIR/radio scatter
and luminosity function to create a candidate scenario for the
CMD relation, as shown in Fig.~5d. 
Scatter in radio/far-IR relation has a skewed effect on the 
data compared to Fig.~5c, in the sense of pulling more high
redshift objects downwards in the diagram.  This can be understood as
due to a selection bias associated with our radio threshold.

This scenario is clearly incompatible with our observed CMD,
failing to reproduce the dearth of sources in the lower right corner of
Fig.~1.  What can we learn from the failure of this picture, to help
understand how to recover the observational trend in the CMD?
Firstly, the model consisted of no luminosity evolution with redshift, and
a mild number density evolution governed by the imposed redshift distribution.
As noted by Blain et al.~(1999b,c), the extremely strong density evolution 
required to fit the 850\mum\ counts (following a $(1+z)^6$ form) will
quickly overproduce the far-IR background.  Boosting our present 
redshift distribution is not a viable modeling route.

The key to the problem lies in
removing sources from the lower right corner, which do not exist at all
in the observations, despite our apparent sensitivity to them. 
As discussed in the previous section, with our toy evolution of a local sample
of {\sl IRAS\/} galaxies, a reasonably strong luminosity evolution on its own
is able to reproduce the general trend of the CMD. 
Several authors have recently modeled the evolution of dusty galaxies
using pure luminosity evolution, 
reproducing the observed counts and backgrounds
at both far-IR and sub-mm wavelengths
(e.g.~Blain et al.~1999b,c; 
Rowan-Robinson 2001; Xu et al.~2001).
The CMD trend can be understood naturally in terms of luminosity evolution:
more luminous objects occur at higher redshifts, thus leaving the
lower right corner largely empty.
We therefore turn to Monte Carlo simulations of
pure luminosity evolution pictures, similar to the models of the
above authors.

We must first consider carefully which local luminosity function
to adopt.  While many authors have adopted a local
{\sl IRAS\/} 60\mum\ luminosity function, Dunne et al.~(2000a) have recently
attempted to construct an 850\mum\ luminosity function directly from
SCUBA measurements of the {\sl IRAS\/}-BGS.
Given that the bright end of the Dunne et al.~function is still
highly uncertain, we follow the lead of the previous modelers and
work from the {\sl IRAS\/} 60\mum\ function as presented in 
Saunders et al.~(1990).  We $K$-correct to 850\mum\ with a single
temperature greybody spectrum to match as closely as possible the
Dunne et al. measurements at 850\mum\ (using $T_{\rm d}\,{=}\,36$,
$\beta\,{=}\,1.3$).

We evolve the local 
luminosity function using $\Phi (L,\nu) = \Phi _0 (L/g(z),\nu (1+z))$.
Our evolution function has a power law peak, $g(z) = (1+z)^{4}$ out
to $z\,{=}\,3$ (the consensus of the median redshift for SCUBA sources
suggested  by Archibald et al.~2001, Smail et al.~2000, Dunlop et al.~2001, 
Scott et al.~2001).  Beyond $z\,{=}\,3$
the function drops again as $g(z) = (1+z)^{-4}$.  This power-law index is
chosen based on evolutionary models fit to both optical and sub-mm wavelength
data (Blain et al.~1999b,c).
To retain the pure luminosity evolution, we truncate the evolved
functions so that they integrate to identical numbers of sources per
comoving volume, as depicted in Fig.~5.

We can then incorporate this luminosity evolution into our Monte Carlo models.
We first introduce our luminosity evolution models into the CMD,
without the radio/far-IR scatter for clarity (Fig.~6a). 
To illustration the
total sample and the effect of radio pre-selection, we show all sources,
but change the symbol size for the sources we can detect (large) and those
that fall below our radio sensitivity (small). Again sources are color-coded
by redshift.  We now find a scenario
which is consistent with the source distribution of our observed CMD,
whilst simultaneously fitting the sub-mm through far-IR counts and backgrounds.
When we bring in the radio/far-IR scatter (Fig.~6b),
we see that the apparent selection bias effect is even larger than in our
previous model (Fig.~4), scattering
sources into the region which lies above the redshift
for which a suitably scaled Arp220 SED would be detectable.

The reasonable match of this model with our measured CMD
additionally suggests that there may be
very high redshift ($z\,{\ga}\,3$) sources in our OFRS SCUBA sample.
However, consideration of the radio undetected sources in the CMD plane 
indicates that the maximum luminosity for which we detect sources in the
sub-mm and radio represents the peak in the adopted evolution function --
sources lying at higher redshifts are necessarily less sub-mm luminous.
Our present model represents the maximum amount of luminosity evolution 
that can be accommodated in such a peak model using the $50\,$K SED template 
without overproducing the 850\mum\ counts (effectively the far-IR
background).
With cooler dust temperatures, less severe luminosity evolution is able to 
reproduce our observed CMD.

\section{Selection function and models}

\subsection{The radio selection function}

Having reproduced our observed CMD data set with a scenario consistent
with galaxy counts at a variety of far-IR and sub-mm wavelengths,
we are now able to better understand our radio selection function
relative to the modeled CMDs. 

We can take our model and derive predicted versus measured counts (Fig.~7)
and redshift distributions (Fig.~8).
With our adopted scatter of 0.2 dex in the farIR/radio correlation, we recover
a large percentage of the sub-mm counts brighter than $10\,$mJy, falling to
a 50\% recovery rate by ${\sim}\,5\,$mJy.
This appears to be less than the recovery rate quoted in BCR and C01. A simple
explanation for this discrepancy is that our Monte Carlo model gives sources
with an infinite effective signal to noise, while the actual data at the
${\sim}\,5\,$mJy level is barely $3\sigma$ significance.  The steep luminosity
function therefore results in a large 
Eddington bias, scattering many ${\sim}\,2\sigma$ sources into our detected
catalog, and making up an apparent excess of source counts in the
${\sim}\,5\,$mJy bin.

Our modeling results have demonstrated that a simplistic estimate of the
radio pre-selection function based on redshifting a single ULIRG falls
short of the true picture.
Considering the entire population of IR-luminous sources, evolving in one
of several plausible scenarios
consistent with our CMD and the 850\mum\ counts, it appears that
higher redshift sources will scatter into our detection window.  Such high-$z$
objects appear to comprise a significant fraction of our brightest sources,
consisting of ${\sim}\,$20\% of the $S_{850}\,{>}\,10\,$mJy detections.
Our analysis of the Monte Carlo
modeled source count and redshift distribution recovery
therefore suggests that we look at direct comparison with the
measured blank field sub-mm properties presented in BCR, C01, and 
Chapman et al.~in preparation.

In Fig.~7b we plot the integrated source counts estimated from our
optically faint radio source sample (filled symbols, with error band).
A fit to the blank field sub-mm survey counts from BCR and Eales et al.~(2000)
are also shown by the dotted line, and the cluster
lensing survey counts of Blain  et al.~(1999a) and Chapman et al.~(2002c)
by crosses and stars, respectively.  The effective
volume of our survey is difficult to estimate, as the radio sensitivity falls
off with radius
from the field center.  This issue has been quantified carefully in C01 (and
will be discussed further in Chapman et al.~in preparation).
Since objects were chosen randomly, we use the visibility function
to construct the count, as discussed in these papers.

The fraction of sources
recovered is clearly high, although the brighter counts are still not
very well constrained.  While the ${\sim}\,2\,$mJy counts
are well determined to be 3000--4000/deg$^2$, brighter than
$5\,$mJy it is not clear how steeply the counts falls.
A recent measurement from Borys et al.~(2001) for sources brighter than
$12\,$mJy in the extended HDF is also plotted, with a square.
This suggests our OFRS recovery rate is about 60\%.

Our modeling analysis lead us to a count recovery rate perhaps as
high as 70\%, in the
absence of large-scale structure effects.  This seems plausible, both from
the direct analysis of the counts comparison (Fig.~7b) and from the sources
which we know to be missing from the OFRS pre-selection: the ${\sim}\,15$\%
optically brighter sources at all redshifts, and the remaining sources at
redshifts too high to be detected in the radio.
This is also in agreement with the direct comparisons of blank field
SCUBA surveys with those sources recovered in the radio from the same fields
(Borys et al.~2001, Smail et al.~2001, C01).
At the  faint end  of the counts (Fig.~7b), our pre-selected sources appear to
fall significantly short of the full source counts.  This was
already evident in our
modeling of the recovery rate (Fig.~7a), where only ${\sim}\,10$\% were
detected at the current radio depths by $S_{850}\,{=}\,1\,$mJy.

In Fig.~8, we focus on the redshift information gleaned from the
850\mum/$1.4\,$GHz
estimators compared to the true distribution.  We take two flux cuts on the
modeled 850\mum\ data and over-plot the true redshift distribution
(green/dashed histogram) on that obtained by directly applying the
850\mum/$1.4\,$GHz redshift 
estimator to the total population (red/solid histogram) and the radio
detected sub-population (blue inset histogram). 
Fig.~8a depicts a $5\,$mJy cut, corresponding roughly to the sensitivity of
our SCUBA OFRS survey, while Fig.~8b presents 
a $2\,$mJy cut, comparable to the deepest observations obtainable with SCUBA.
Clearly in the absence of 850\mum/$1.4\,$GHz scatter,
we recover the true redshift distribution.
Introducing the scatter results in a broadening and skewing of
the distribution, as a result of the strong luminosity evolution
adopted in our model.
The radio selection (blue inset histogram) results in a
bias, whereby objects at higher redshift, beyond the peak in our evolution
function $g(z)$, 
are scattered into our radio detection window.  However,
these appear as an excess of $z\,{\sim}\,2$ sources,
having been scattered down from the large population with ever decreasing
characteristic luminosity ($L^*$ in our luminosity functions). 
The excess of measured high-$z$ sources arises
simply from the intrinsic scatter in the 850\mum/$1.4\,$GHz relation,
broadening the peak in our evolution function, $g(z)$.
These features must be ubiquitous in any realistic evolutionary model
which rises in luminosity to a peak redshift and then declines to even
higher redshifts.

These biases are quantified by comparing the actual
redshift distribution recovered by the radio selection (blue histogram) to
the recovery in the absence of scatter in the far-IR/radio relation
(Figs.~8c,d).
In our $5\,$mJy flux cut, we recover an excess of sources for all $z\,{>}\,3$
but recover a deficit of sources from $2\,{<}\,z\,{<}\,3$.
Our radio selected survey
is therefore less complete at  $2\,{<}\,z\,{<}\,3$ than we would have naively
expected based on the redshift at which Arp220 is no longer detectable in
the radio.  This is important for spectroscopic follow-up of these samples
(Barger et al.~1999, Ivison et al.~2000, Chapman et al.~2002b).
In Figs.~8e,f, we also plot the residuals, as an estimate of the errors
involved in both the intrinsic radio selection and the 850\mum/$1.4\,$GHz
estimator. 
The dashed line (green) shows the residuals of those radio detected sources
with an 850\mum/$1.4\,$GHz estimated redshift distribution compared
with the true total distribution. 
The solid line (yellow) shows the true redshifts of the radio selected
sources from the total distribution.  Finally the light solid line (blue)
shows the difference in the 850\mum/$1.4\,$GHz estimated redshift from the
true redshift for the radio selected sources alone.

In Fig.~9, we plot the 850\mum/$1.4\,$GHz estimated redshift distribution for
the actual SCUBA observed sources in our sample, using the same Arp220
template SED used in the models to derive redshifts.  Utilising a template with
cooler dust would of course result in lower redshift estimates.

\subsection{The optical selection function}

The optical selection function rejects any optically brighter
sub-mm sources, which comprise ${>}\,10$\% of the Smail et al.~(2000)
sub-mm survey.  A constant optical limit ($I\,{\ga}\,24$) in the
absence of evolution would select more obscured sources at lower redshifts.
However, Adelberger \& Steidel (2000)
and Chapman et al.~(2002d) have suggested that star-bursting sources
appear less obscured at higher redshifts, counteracting the effect of
cosmological dimming.  In fact a constant optical
limit may produce the most uniform sample of sources across the
$z\,{=}\,1$--3 range.

The majority of the high redshift ULIRG population, as detected by SCUBA,
has been demonstrated to be optically faint, typically with $I\,{>}\,24$,
$K\,{>}\,20$ (BCR; C01; Smail et al.~2001; Chapman et al.~2002a). 
Obtaining
their redshift distribution is crucial for constraining their
evolution and most importantly for placing them in their cosmological context
amongst the hegemony of other galactic denizens (e.g.~Blain et al.~1999b,c).
With typical magnitude of $I\,{>}\,24$, however, present generation telescopes
are hard pressed to routinely measure redshifts for such distant, faint and
highly obscured galaxies.  This is particularly disappointing given that
accurate redshifts are essential for obtaining CO observations of these
gas-rich galaxies.  The OVRO observations presented by
Frayer et al.~(1998, 1999) and Chapman et al.~(2002e -- 
the SA22~Ly$\alpha$\,blob) of
three SCUBA galaxies demonstrate the important physical information that can
be obtained if accurate redshifts are available.

Fortunately, around 10\% of the SCUBA-detected galaxies are unambiguously
identified with relatively
bright ($I\,{<}\,24$) counterparts (the class 2 sources in the nomenclature of
Ivison et al.~2000, a class which includes the two systems targeted by
Frayer et al.).  Such galaxies are spectroscopically observable with
10m-class telescopes and so could offer one route to estimating the
redshift distribution of the population as a whole.  However, we first
have to address the question of the relevance of these optically brighter
cousins to the bulk of the SCUBA population with optically faint
counterparts -- are the class 2 galaxies representative of the whole or are
they an unrelated population, or perhaps a different episode in the
evolutionary sequence?

We have shown some evidence
that in fact the optically `brighter' sub-mm sources may in fact
be representative of the larger optically faint population (Fig.~1) --
the CMD reveals that
optically brighter galaxies are scattered uniformly amongst the fainter
population.

At least in part this effect may be explained because some systems may
comprise both obscured and unobscured interacting components,
e.g.~SMM14011 from Ivison et al.~(2000) and MMD11 from Chapman
et al.~(2000,2002f)  -- merging systems with the bulk of the
sub-mm emission arising in one galaxy component, while the optical
emission remains unobscured in an associated component.  Equivalently the
two `components' may be mixed up within a single galaxy, through patchy
dust and star-formation.  Our
radio-selection of optically faint sub-mm sources is blind to these
possibilities, and optically brighter sources are often located very nearby
the radio position.  In other cases the galaxy may simply have substantial
outflows which have expelled dust along prefered viewing directions,
allowing a view through the veil.

\section{The 450\mum\ detections and the color-color relation}

Many of our sources are also detected at 450\mum, which allows for the 
possibility of improving the constraints on evolutionary properties for the
sample.  We have extracted 450\mum\ flux densities from the models of
Fig.~6 and and plotted CMD and color-color diagrams in
Fig.~10, along with the 850\mum/radio CMD (reproduced without the
confusion of sources falling beneath our radio limit).  In this figure the
the redshifts of the sources are mapped onto a continuous color code,
represented by the bar in Fig.~10.
 
While some of the sources are only marginally detected
at 450\mum, the
firm detections obtained at 850\mum\ and at $1.4\,$GHz lend support to their
reality.  We focus on the 19
sources with 450\mum\ flux detected at greater than 2$\sigma$.
These data are presented in Fig.~11, where we plot
both a 450\mum/850\mum\ CMD and a 450\mum/850\mum\ versus 850\mum/$1.4\,$GHz
CCD.  The envelopes of our Monte Carlo models are shown as dashed lines.
For comparison, we also plot the 5 sources from the Smail et al.~(2001) sample 
detected at 450\mum, and the single source from Eales et al.~(2000) by
triangles and squares, respectively.
The average (inverse variance weighted) signal at
450\mum\ of the 850\mum\ detected sources is $34\pm6\,$mJy,
consistent with a population with median $z\,{=}\,3$ and
our adopted dust parameters.

An additional consideration for 450\mum\ observations is the 
increased instability in the sky opacity at this shorter wavelength, and
the larger effect of extinction corrections as a 
consequence, leading to possibly larger flux uncertainties than naively
estimated.  The importance of such effects was discussed recently by
Dunne \& Eales (2001) for a sample of local galaxies measured at 450\mum.
We obtained calibration maps for our
data at 6 hour intervals on average, and the data were all taken near the
middle of the night where the measured gains at 450\mum\ are far more
stable than at the extremities of the night.  While the absolute calibration
is therefore not necessarily the dominant error on our measured flux densities,
the current inability to measure atmospheric opacity rapidly
during observations precludes calibrating the data to better than 20\%, and we
adopt this as our nominal calibration error.

Recall that locally (Section~4), the luminous IR sources have a relatively
tight 450\mum/850\mum\ correlation, but a much larger scatter in
100\mum/850\mum\ and 60\mum/850\mum, 
the latter two being closer in rest wavelength to 450\mum\ in
our highest redshift SCUBA sources.
Therefore, the relatively flat relation of 450\mum/850\mum\
with 850\mum\ flux observed for our OFRS SCUBA observations
is suggestive of a sample which is relatively uniform in dust properties.
The Smail et al.~(2001) sources have amongst the lowest 450\mum/850\mum\
ratios, which could be
consistent with redshift effects dominating their sub-mm ratios.
Our selection bias to lower redshift sources is much stronger than in
their blank field survey due to the negative sub-mm $K$-correction.
However, the large error bar on our measurements, and the associated strong
Eddington (noise) bias from selecting all sources with $S_{450}\,{>}\,2\sigma$
is also likely to be a factor.
The correlation of 850\mum to $1.4\,$GHz fluxes, coupled with the skewed
detection of higher redshift sources from the Eddington bias,
results in an even tighter relation in the 450\mum/850\mum\ versus
850\mum/$1.4\,$GHz color-color space.

If the higher redshift sources are the most luminous, as suggested by the
correlation in 850\mum\ versus $1.4\,$GHz fluxes,
we expect a smaller 450\mum/850\mum\ with
increasing 850\mum\ flux, a result of the increasing redshift.
However, we also expect brighter sources to have hotter dust, based
on the local {\sl IRAS\/} correlation, resulting in a larger 450\mum/850\mum\
ratio with increasing 850\mum\ flux.
These two effects will therefore partially cancel in the 450\mum/850\mum\
versus 850\mum\ CMD, resulting in a flat relation.

A scenario with no variation in dust properties or luminosity evolution
could also produce such a flat relation.  The spread in 450\mum/850\mum\
would then be simply a result of the redshift distribution of objects with
identical dust properties.
However, the radio/sub-mm correlation introduces either luminosity evolution or
dust correlation with luminosity.
The flat relation must therefore be the result of some intrinsic property.

Our current data do not allow robust constraints on  
situations where either the dust properties evolve with
redshift (e.g.~temperature increasing with redshift),
or the population has a larger spread in dust properties
with increasing redshift.  While the conditions were generally
rather poor at 450\mum\ during our JCMT runs, the 850\mum\ detected
sources are mostly luminous enough to be detectable with SCUBA at
450\mum\ under more favorable conditions.  Better 450\mum\ data would
certainly improve the constraints on models.

%
\section{Discussion}

\subsection{Density evolution and the nature of sub-mm galaxies}

Studies of hierarchical galaxy formation pictures suggest that 
significant density evolution must occur (e.g.~Cole et al.~2000, 
Blain et al.~1999c). 
Successful multi-wavelength evolutionary scenarios
have already been explored within this paradigm (e.g.~Blain et al.~1999c).
However the question for our purposes concerns the significance of
merging sub-fragments for the
15 arcsec scales probed by the SCUBA beam.
The SCUBA sources seem likely to trace the merger events themselves rather
than the fragments coming together before merging (and perhaps surviving
afterwards without any significant far-IR emission).
Therefore pure luminosity evolution is not necessarily a poor model for
the bright end of the SCUBA counts.
The situation described above for optically bright versus faint SCUBA sources
is also relevant here -- 
luminous sources may sometimes pair with a merging optically bright source,
or a snapshot in the merger sequence may have optically luminous and
dust-enshrouded parts.

Number evolution, however, is not only driven by merging. 
The SCUBA-bright population (if driven by star formation) may represent a 
transitory stage in a galaxy's evolution, possibly related to 
the AGN phenomenon (e.g.~Lewis \& Chapman~2002, Archibald et al.~2001).

The effect of currently fashionable flat, $\Lambda\,{\simeq}\,0.7$
cosmologies is to make the situation worse by introducing a
slope to the color-magnitude plot, which also has to be countered with
luminosity evolution.  This would require even stronger evolution than the 
$(1+z)^4$ adopted in our present study, assuming the characteristic
dust temperature of the sources is unchanged.
An alternative would be to bring in an additional density evolution
component.  We have certainly not demonstrated that our particular model
is a unque solution.  But it seems that strong luminostity evolution is
a requirement of an successful model.

\subsection{The nature of the sources and future work}

The fainter sub-mm sources, marginally detected at 2--$6\,$mJy,
clearly follow the observed and modeled CMD trend.
This suggests that pursuing these candidate fainter sources
with deeper SCUBA integrations could yield significant detections,
providing a valuable lever arm in luminosity for modeling the evolutionary
properties of the sample.
The most luminous $z\,{\sim}\,3$ Lyman-break galaxies (LBGs) are known to
emit at the 1--$2\,$mJy level (Chapman et al.~2000,2002d;
Peacock et al.~2000), so their $z\,{=}\,1$--2 counterparts
probably contribute some fraction of these fainter sub-mm sources.
This is consistent
with the high SFR deduced by Steidel et al.~(1999), who apply a large
dust correction to their results.

Note that whether or not a source has an AGN may not have a tremendous
influence on its position in the CMD, especially if the energetics are
still dominated by star formation.  It has also been pointed out many times
(e.g.~Sanders 2001)
that AGN and star-bursts are both related to merging, and so AGN and
star-bursts could be the same objects caught at different evolutionary
phases.  In that case many of the correlations between fluxes at different
wavelengths may continue to hold, at least for weak AGN.

Ho et al.~(2000) have investigated the
strength of the radio synchrotron emission from Seyferts and LINERs locally.
The radio-loud AGN effect on the CMD is to lower points (brighten the radio),
which cannot easily be compensated for by other properties,
dust temperature for example.  This suggests further that it is unlikely
that radio-loud AGN dominate the sub-mm luminous sources.
One source from Smail et al.~(2001), which has been identified with an AGN 
component
inhabits the lower cut-off region of the CMD, suggesting that this region may
in general be inhabited by those sub-mm sources with stronger AGN components
(a stronger AGN radio source).
However, SMM14011 (Ivison et al.~2000), which has
been confirmed spectroscopically as a star-forming galaxy, also inhabits this
region, and we can draw no firm conclusions that AGN show their presense in
any obvious manner in our CMD or color-color diagnostics.

Future work identifying the actual counterparts at multiple
optical wavelengths, as well as obtaining a larger sample of 
radio sources with SCUBA measurements (both optically bright and faint)
will lead to substantial refinement of our understanding of these issues.
Higher spatial resolution HST observations may
also elucidate the morphologies of the sources, allowing an understanding
of how the radio/sub-mm emission is traced by the optical, and where the 
heart of the far-IR energy is generated in high redshift objects.
Deeper sub-mm observations of the sub-mm faint radio sources will
allow us to place them in context relative to the sub-mm
bright radio sources. 

%
%
\section{Conclusions}

We have presented a large sample (55) of sub-mm sources with radio and
optical counterparts.
We observe a correlation between the sub-mm flux and the sub-mm/radio 
which implies that higher redshift sources must be more luminous.
Local correlations in the radio/sub-mm/far-IR properties of
luminous infrared galaxies provide a benchmark for assessing the trends
observed in our higher redshift analogs.  In particular, the effects of dust
temperature on our evolutionary scenario can be directly assessed.
Using a Monte Carlo approach to model the evolution of the IR
luminous galaxies, we are able to simultaneously
represent the CMD, counts and sub-mm/radio implied redshift 
distribution for our observed sample.  Our models are not unique, but
indicate some general features which any successful model must possess,
in particular this strong luminosity evolution.

The selection function from sub-mm follow-up to 
optically faint radio can be better understood in relation to our Monte
Carlo models.
There are inherent biases in the technique beyond that expected from a 
naive redshifting of the Arp220 SED out to redshifts where it is no longer
detectable in the radio. 
The combination of radio/far-IR scatter and the necessary rise and fall in
the evolution function results in a bias in the selection of sources 
(if we were able to 
obtain spectroscopic redshifts for the entire sample).  We uncover an
excess of high redshift sources and lose some of the lower redshift
sources that would otherwise be present in our selection window. 
In the absence of spectroscopic redshifts, the reconstruction of the redshift 
distribution is broadened and weighted towards lower redshifts.
The few optically bright sub-mm sources in our sample are distributed
uniformly throughout the CMD with no prefered locus.  This suggests that
spectroscopic follow-up of the
optically brighter population may provide a reasonable approximation to the
true redshift distribution.
	
This redshift distortion presents the 
possibility of a population of very high
redshift ($z\,{>}\,3$) star forming galaxies being detected with our radio
selection approach.
With current radio flux limits, the required luminosities are too large
to be included in our sample for $z\,{>}\,5$ under the assumptions of our
present model.

A subsample of 25 sources in our sample are marginally detected in the
450\mum\ band. When compared to local {\sl IRAS\/}
galaxies, these suggest a similar 
range in dust properties to local ULIRGs, possibly with somewhat hotter
or more emissive dust.

\section*{Acknowledgements}

We thank the  staff of the JCMT, 
for their assistance with the observations. 
The referee, C.Carilli, has helped us to improve the final version
of this paper.
A special thanks to R.~Ivison, I.~Smail, A.~Blain, and  
L.~Dunne for helpful discussions on this work.
We gratefully acknowledge support from NASA through
HST grant 9174.1 (SCC) and
Hubble Fellowship grant HF-01117.01-A (EAR) awarded by the
Space Telescope Science Institute, which is operated by the
Association of Universities for Research in Astronomy, Inc.,
for NASA under contract NAS 5-26555.
GFL thanks the Australian Nuclear 
Science and Technology Organization (ANSTO) for financial support.
DS and CB are supported by the Natural Sciences and Engineering Research
Council of Canada.

\section*{references}

\noindent  Adelberger, K., Steidel, C., 2000, AJ, 119, 2556

\noindent  Archibald, E., et al., 2001, MNRAS, 323, 417

\noindent  Barger., A., Cowie, L., Mushotzsky, R., Richards, E., 2001a, 
	AJ, 121, 662

\noindent  Barger, A., et al., 2001b, ApJ, in preparation 

\noindent  Barger, A., et al., 2001c, ApJ, in press 

\noindent  Barger, A., Cowie, L., Richards, E., 2000a, AJ 119, 2092 (BCR)

\noindent  Barger, A., et al., 1999, AJ 117, 2656  

\noindent  Barger, A., et al., Nature, 1998, 394, 293  

\noindent  Bertoldi, F., et al., 2000, A\&A, 360, 92

\noindent  Blain, A., 2000, MNRAS, 309, 955 

\noindent  Blain, A., et al., 1999a, ApJL, 512, 87 

\noindent  Blain, A., et al., 1999b, MNRAS 302, 632 

\noindent  Blain, A., et al., 1999c, MNRAS, 309, 715 

\noindent  Blain, A., 1999, MNRAS, 309, 955 

\noindent  Borys, C., Chapman, S.C., Scott, D., Halpern, M., 2001, 
	MNRAS, in press

\noindent  Borys, C., Chapman, S.C., Scott, D., 1998, MNRAS 308, 527 

\noindent  Carilli, C., Yun, M., 2000, ApJ 530, 618

\noindent  Carilli, C., Yun, M., 1999, ApJ 5130, 13L

\noindent  Chapman, S.C., Richards, E., Lewis, G.F., Wilson, G., Barger, A.,
	2001a, ApJL, 548, 147 (C01)


\noindent  Chapman, S.C., et al. 2002a, ApJ, in preparation 

\noindent  Chapman, S.C., et al. 2002b, ApJL, in preparation 

\noindent  Chapman, S.C., Scott, D., Borys, C., Fahlman, G., 2002c, MNRAS, in press  

\noindent  Chapman, S.C., et al. 2002d, MNRAS, submitted  

\noindent  Chapman, S.C., et al. 2002e, ApJL, in preparation 

\noindent  Chapman, S.C., et al. 2002f, ApJL, submitted 

\noindent  Chapman, S., Scott, D., Steidel, C., et al., 2000, MNRAS, 319, 318

\noindent  Cole, S., et al., 2000, MNRAS, 319, 168

\noindent  Connolly, A., et al., 1997, ApJ 486, L11 

\noindent  Condon, J., Helou G., Sanders D., Soifer, B., 1996, ApJS, 103, 81 

\noindent  Condon, J., 1992, ARA\&A, 30, 575

\noindent  Dale, D., et al., 2001a, ApJ, 549, 215

\noindent  Dale, D., et al., 2001b, ApJ, submitted


\noindent  Dunlop, J., 2001, 
FIRSED2000 conference proceedings, eds. I.M. van Bemmel, B. Wilkes,
    \& P. Barthel, Elsevier New Astronomy Reviews

\noindent  Dunne, L., et al., 2000a, MNRAS, 319, 813 

\noindent  Dunne, L., et al., 2000b, MNRAS, 315, 115 

\noindent  Dunne, L., Eales, S., 2001, MNRAS, in press 

\noindent  Eales, S., et al., 2000, AJ, 120, 2244



\noindent Fixsen, D.J., et al., 1998, ApJ, 508, 123

\noindent Fomalont, E., et al., 2002, ApJ, in press

\noindent Frayer, D., et al., 1999, ApJ 514, L13 

\noindent Frayer, D., et al., 1998, ApJ 506, L7 


\noindent  Hauser, M., et al., 1998, ApJ, 508, 25

\noindent  Helou, P., et al., 1985, ApJ 440, 35 

\noindent  Ho, L., Peng, C.Y., 2001, ApJ, 555, 650

\noindent  Holland, W., et al., 1999, MNRAS 303, 659 




 
\noindent  Ivison, R., et al., 2000, MNRAS, 315, 209  



\noindent  Lewis, G., Chapman, S.C., 2000, MNRAS, 318, 31L

\noindent  Lewis, G., Chapman, S.C., 2002, MNRAS, submitted



\noindent  Peacock, J., et al., 2000, MNRAS, 318, 535

\noindent  Puget, J.-L., et al., 1996, A\&A 308, L5 



\noindent  Richards, E., 2000, ApJ 533, 611

\noindent  Richards, E., et al., 1999, ApJ 526, 73L

\noindent  Rowan-Robinson, M., 2001, ApJ, 549, 745

\noindent  Sanders, D.B., 2001, `QSO Hosts and their Environments', in press,
astro-ph/0109138

\noindent  Saunders, R., et al., 1990, MNRAS 257, 425 

\noindent  Scott, D., et al., 2000, A\&A 357, L5 

\noindent  Scott, S., et al., 2001, submitted

\noindent  Smail, I., et al., 2001, ApJ in press

\noindent  Smail, I., et al., 2000, ApJ 528, 612 

\noindent  Smail, I., et al., 1999, MNRAS, 308, 1061 

\noindent  Smail, I., Ivison, R.J., Blain, A.W., 1997, ApJ 490, L5

\noindent  Steidel, C., Adelberger, K., Shapley, A., Dickinson, M.,
 Pettini, M., Giavalisco, M., 2000, ApJ, 532, 170

\noindent  Steidel, C., Adelberger, K., Giavalisco, M., Dickinson, M.,
 Pettini, M., 1999, ApJ 519, 1 

\noindent  Xu, C., et al., 2001, ApJ, submitted

\noindent  Yun, M., Reddy, N., Condon, J., 2001, ApJ, 554, 803

\clearpage


%
%
%
\begin{figure*}
\begin{center}
\epsfig{file=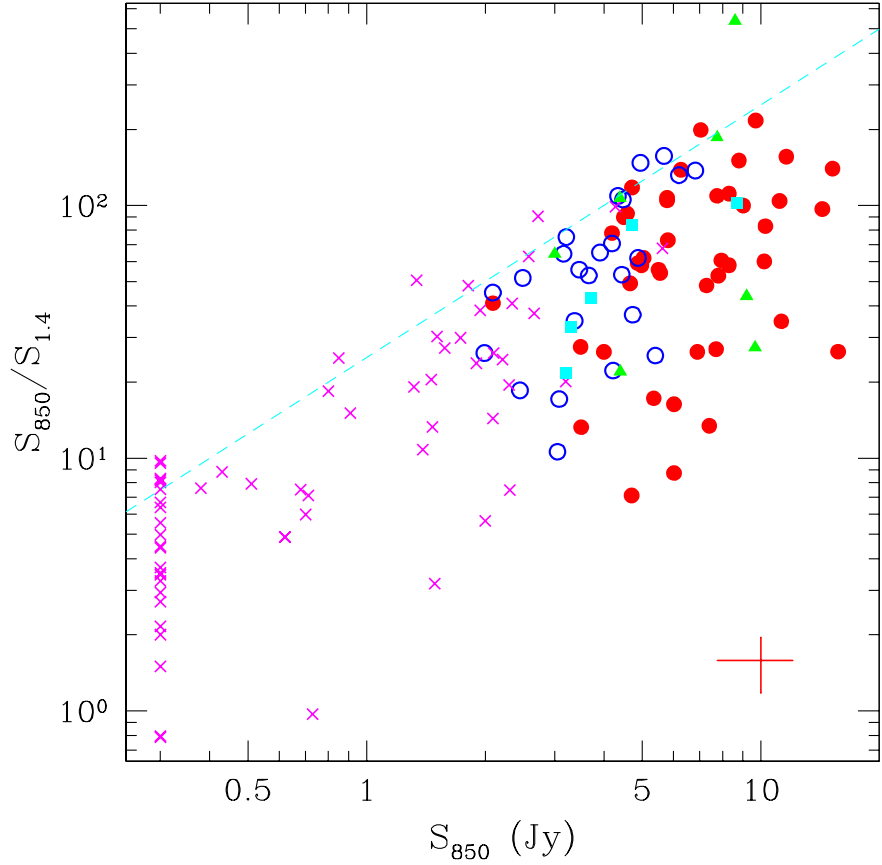,height=12.25cm,angle=0}
\end{center}
\label{X}
\figcaption[chapman.fig133]{
The radio, $S_{850 \mu m}$ color-magnitude diagram.
Red circles are radio sources detected with ${\ga}\,3\sigma$ significance at
850\mum, while blue circles are an additional sample of sources marginally
detected at $>2\sigma$.   The error bar n the lower right is representative
for the detected sample.
Crosses show radio sources with ${<}\,2\sigma$ significance at 850\mum,
where those with observationally estimated fluxes which are negative being
plotted at $S_{850}=0.3$\,mJy.
We also plot the sub-mm sources with radio counterparts from Smail et
al.~(2001) and Eales et al.~(2000) as 7 triangles and 5 squares respectively.
The typical radio flux limit of the sample is shown as a dashed line;
the two sources lying above the line were selected from the deeper SSA13
radio field.
A correlation is apparent in the data, but this is largely defined by
the upper envelope of the radio sensitivity cut-off. 
}
\end{figure*}

%
%
%
\begin{figure*}
\begin{minipage}{170mm}
\begin{center}
\epsfig{file=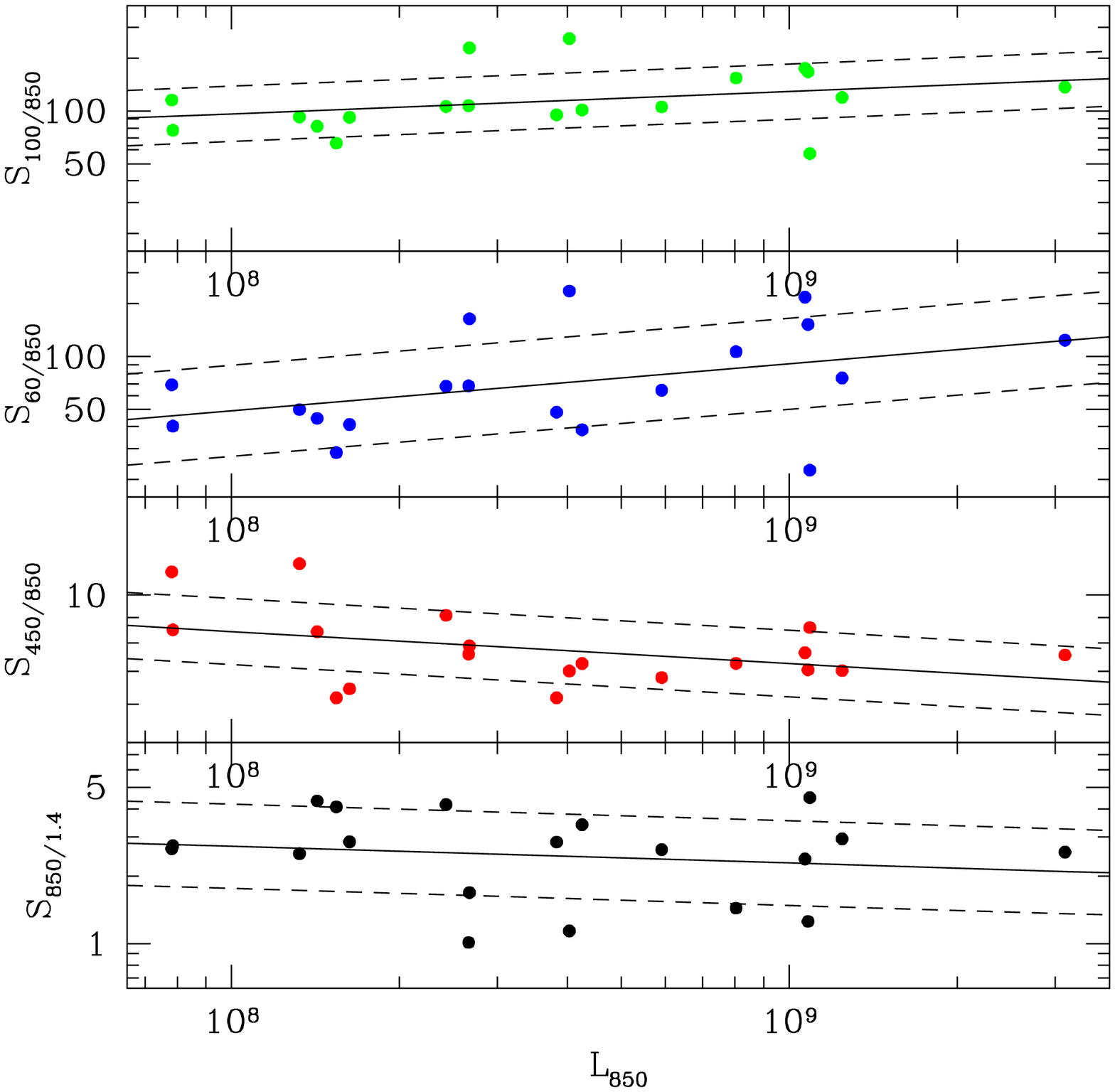,height=8.25cm,angle=0}
\epsfig{file=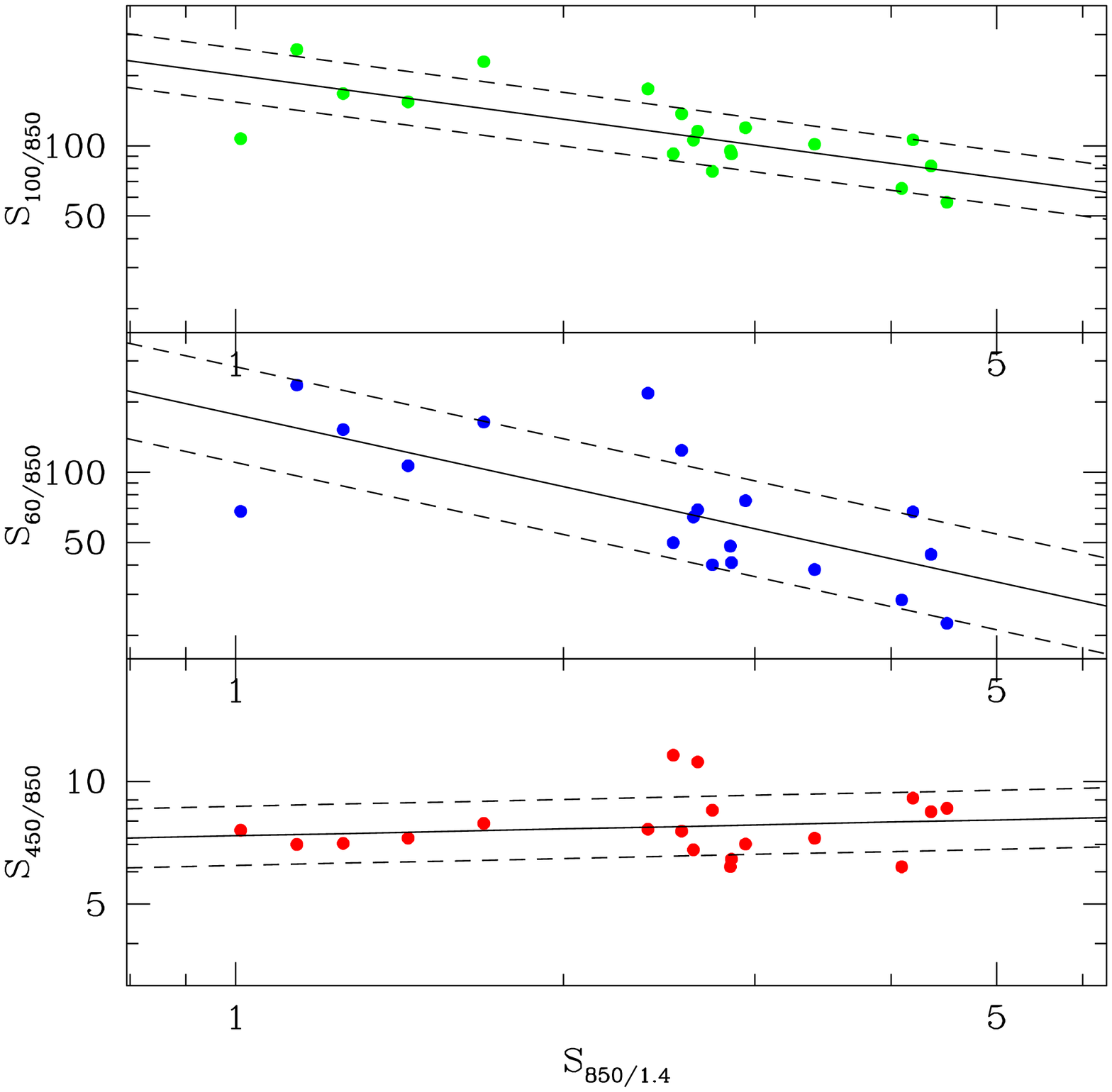,height=8.25cm,angle=0}
\epsfig{file=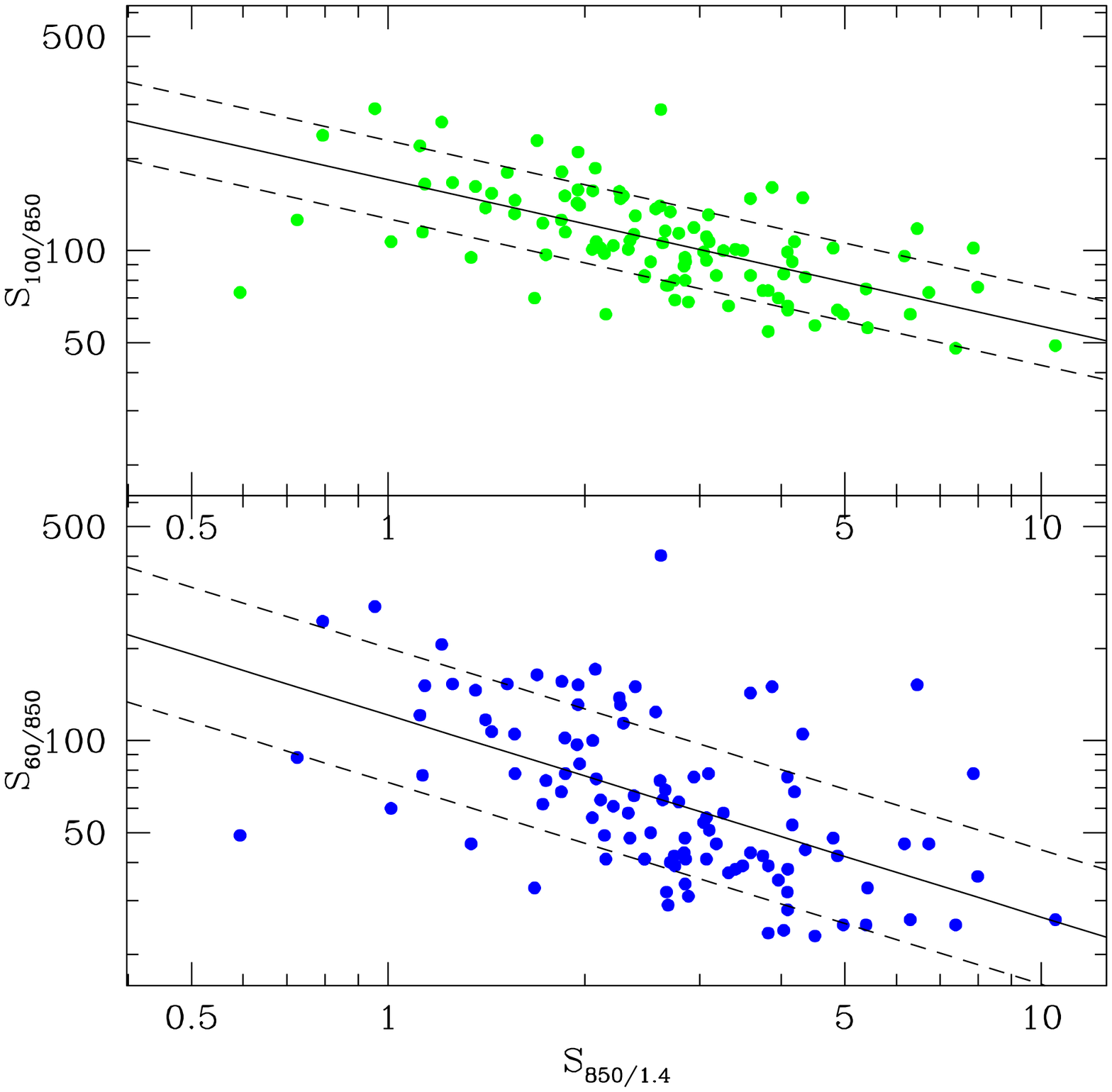,height=8.25cm,angle=0}
\end{center}
\label{X}
\figcaption[chapman.fig3]{Local CMD and color-color relations for
{\sl IRAS\/} luminous galaxies.
{\bf Upper left:} CMDs for 19 galaxies observed at 850\mum\ and 450\mum\
by Dunne \& Eales (2001), with L$_{850}$ in units of L$_\odot$.  
Dashed lines represent the scatter about the
best-fit relations (solid lines).
{\bf Upper right:} Color-color diagrams (CCDs) for these same 19 galaxies.
Rather steep power law correlations are observed in the 100\mum/850\mum\
and 60\mum/850\mum\ versus 850\mum/$1.4\,$GHz
diagrams (with indices -0.63 and -1.03 respectively), while
the 450\mum/850\mum\ relation remains flat. 
{\bf Lower}
Color-color diagrams for the larger {\sl IRAS\/} sample from Dunne
et al.~(2000a)
having only 850\mum\ follow-up.
}
\end{minipage}
\end{figure*}

%
%
\begin{deluxetable}{lrrr}
\tablewidth{480pt}
\scriptsize
\tablenum{1}
\label{table-1}
\tablecaption{\sc \small Local {\sl IRAS\/} galaxy radio/sub-mm/far-IR correlations}
\smallskip
\tablehead{
\colhead{relation} 
 & \colhead{slope$^a$} &  \colhead{log($y$-intercept)$^b$} &
 \colhead{scatter (dex)$^c$}\\ } 
\startdata
100\mum/850\mum\ vs 850\mum & 0.13 & 0.97 & 0.16  \\
60\mum/850\mum\ vs 850\mum & 0.27 & $-0.45$ & 0.26  \\
450\mum/850\mum\ vs 850\mum\ & $-0.07$ & 1.45 & 0.07  \\
\smallskip
850\mum/$1.4\,$GHz vs 850\mum & $-0.08$ & 1.04 & 0.19  \\
100\mum/850\mum\ vs 850\mum/$1.4\,$GHz & $-0.63$ &2.30 &0.12  \\
60\mum/850\mum\ vs 850\mum/$1.4\,$GHz & $-1.03$ &2.25 &0.21  \\
\smallskip
450\mum/850\mum\ vs 850\mum/$1.4\,$GHz & 0.06 &0.87 &0.07  \\
100\mum/850\mum\ vs 850\mum/$1.4\,$GHz$^d$ & $-0.48$ & 2.23 & 0.13  \\
60\mum/850\mum\ vs 850\mum/$1.4\,$GHz$^d$ & $-0.66$ & 2.08 & 0.22  \\
\enddata
\vspace*{-0.5cm}
\tablerefs{Data taken from Dunne et al.~(2000a) and Dunne \& Eales (2001).\\
$^a)$ power law slope index $S_\nu \propto \nu^\alpha$.\\
$^b)$ the log of the power law coefficient $A$ in 
 $S_\nu = A \nu^\alpha$\\
$^c)$ scatter in power law fit to the data measured in dex\\
$^d)$ fits to 121 galaxies having only 850\mum\ flux densities
from Dunne et al.~(2000a) rather than the 19
from Dunne \& Eales (2001) with additional 450\mum\ measurements.}
\end{deluxetable}

%
%
%
\begin{figure*}
\begin{minipage}{170mm}
\begin{center}
\epsfig{file=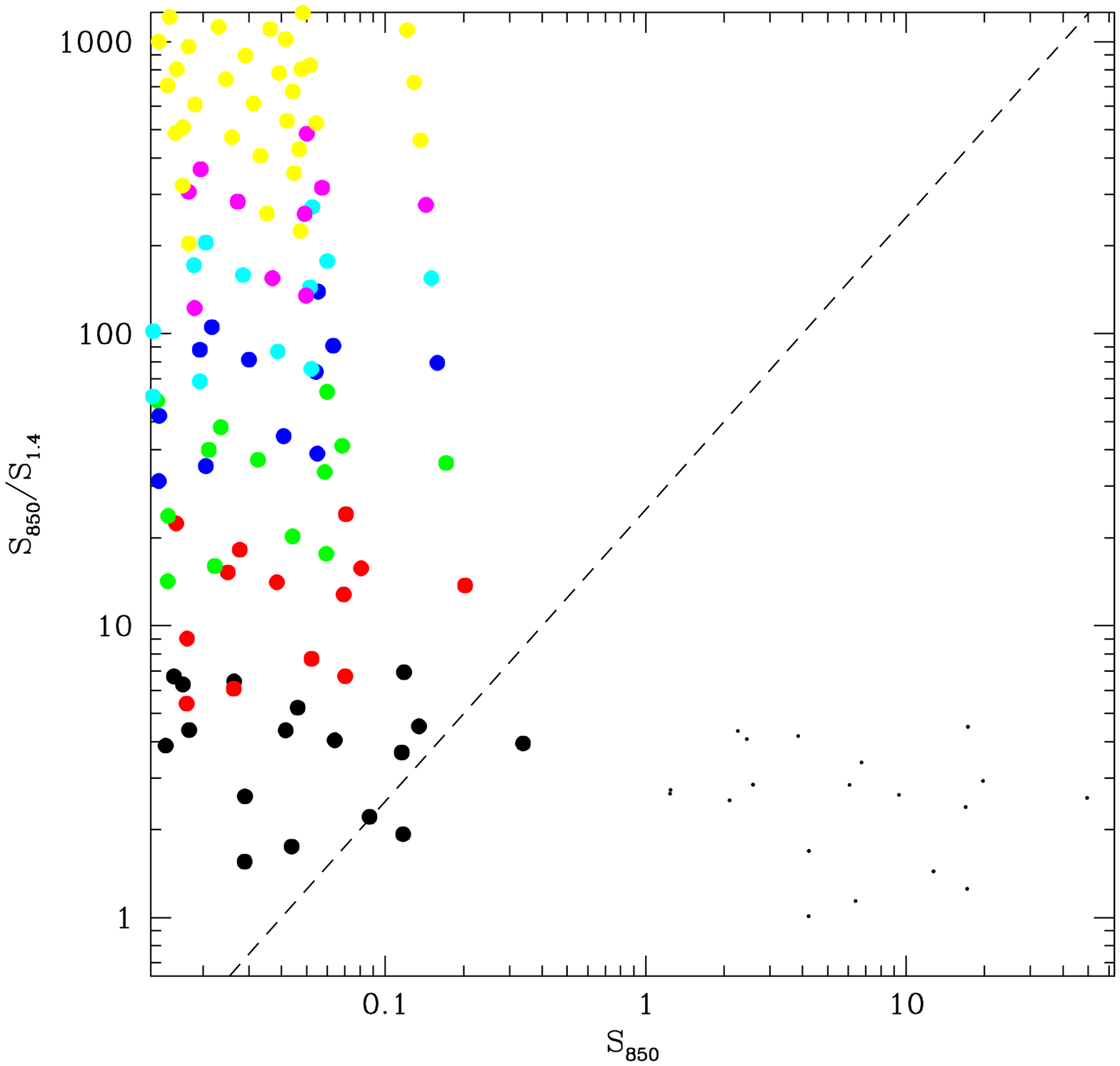,height=8.25cm,angle=0}
\epsfig{file=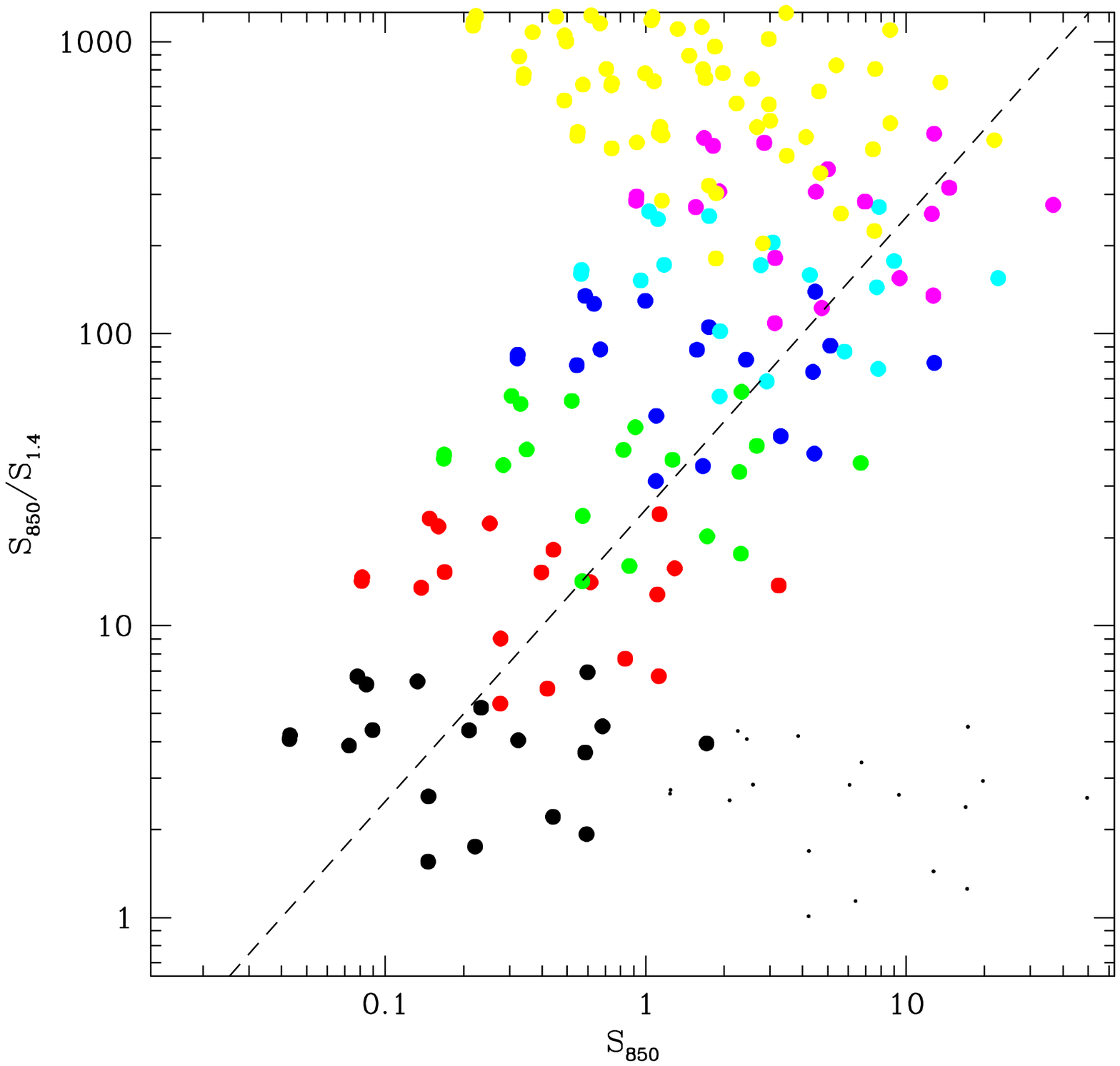,height=8.25cm,angle=0}
\epsfig{file=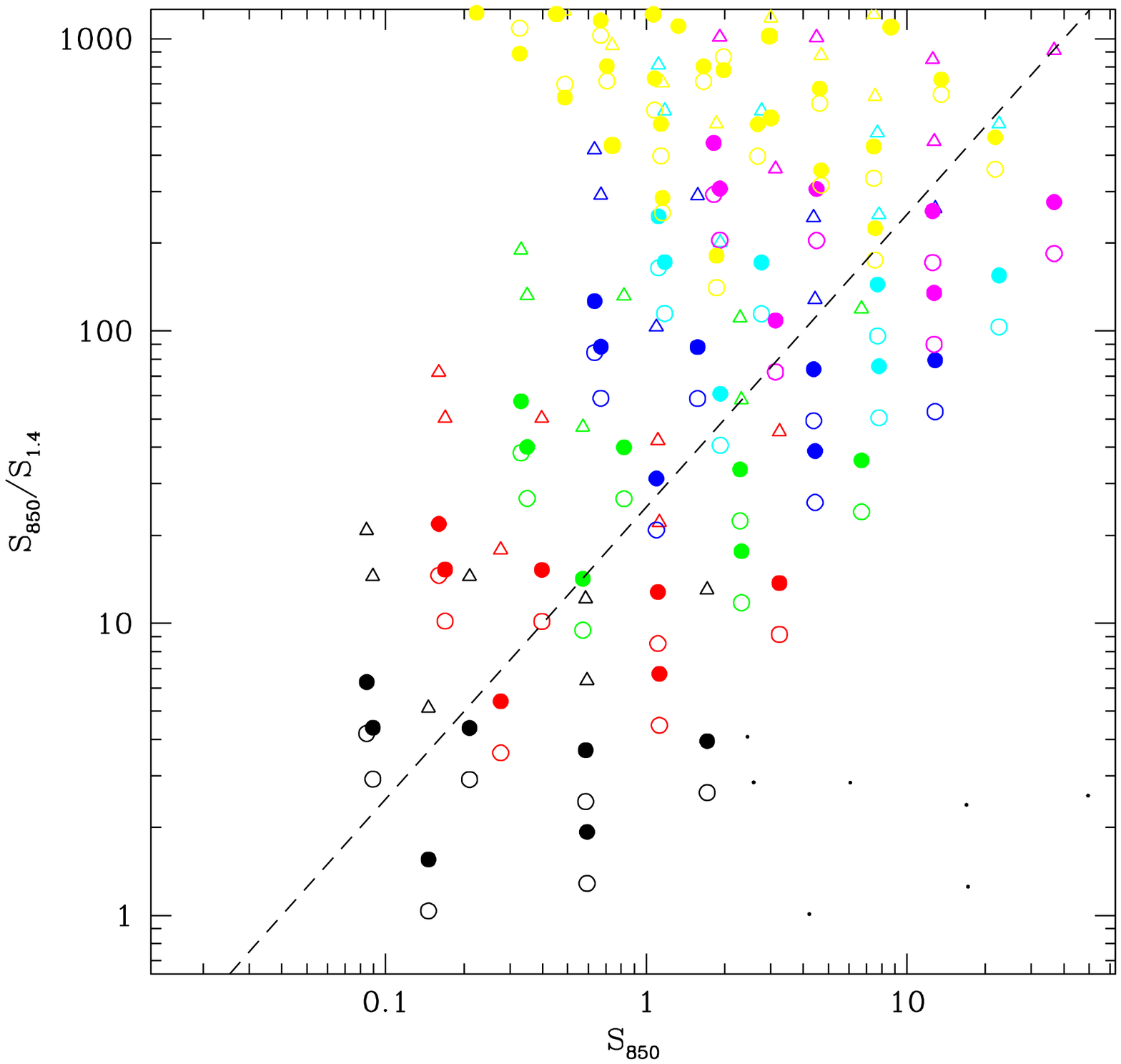,height=8.25cm,angle=0}
\end{center}
\label{X}
\figcaption[chapman.fig19]{
{\bf Upper left:}
The 19 {\sl IRAS\/}
sources from Dunne \& Eales (2001), shown at $z\,{=}\,0$ by the points,
have been redshifted, as indicated by the circles.  Flux densities are
calculated using the locally estimated luminosties using an
$\Omega_{\rm M}\,{=}\,1.0$ cosmology.  Beyond $z\,{\simeq}\,1$ the sources
retain approximately the same observed flux density, whilst increasing in 
850\mum/radio as a function of redshift (i.e.~moving vertically on the plot).
{\bf Upper right:}
Evolution of the galaxies in luminosity, scaling by $(1+z)^4$ out to
$z\,{=}\,3$, in an attempt
to match our observed CMD from Fig.~1.  Conversion from luminosity to
flux density space, results in the sources traveling roughly along the 
suggested CMD correlation,
but then turning above our radio limit (dashed line) beyond the peak
in the adopted luminosity evolution at $z\,{=}\,3$.
{\bf Lower panel:}
Evolution of a subset of the local sources with cooler
($T_{\rm d}\,{=}\,25$\,K -- triangles), medium ($T_{\rm d}\,{=}\,36$\,K
-- open circles: the mean dust temperature from Dunne et al.~2000a) and 
hotter ($T_{\rm d}\,{=}\,45$\,K -- open circles)
dust temperatures.  The 850\mum\ luminosity is kept fixed here. 
}
\end{minipage}
\end{figure*}

%
%
%
\begin{figure*}
\begin{minipage}{170mm}
\begin{center}
\epsfig{file=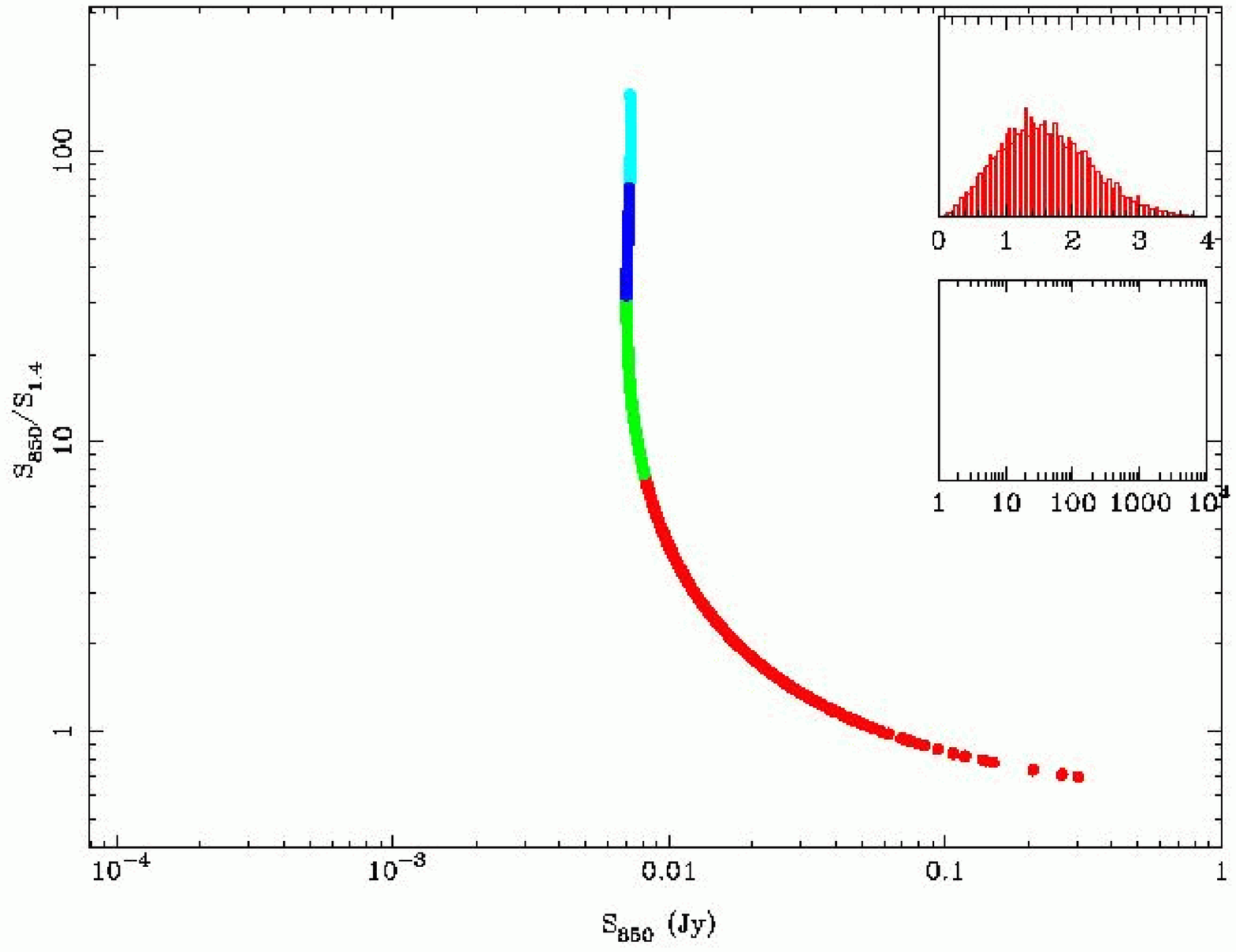,height=6.25cm,angle=-90}
\epsfig{file=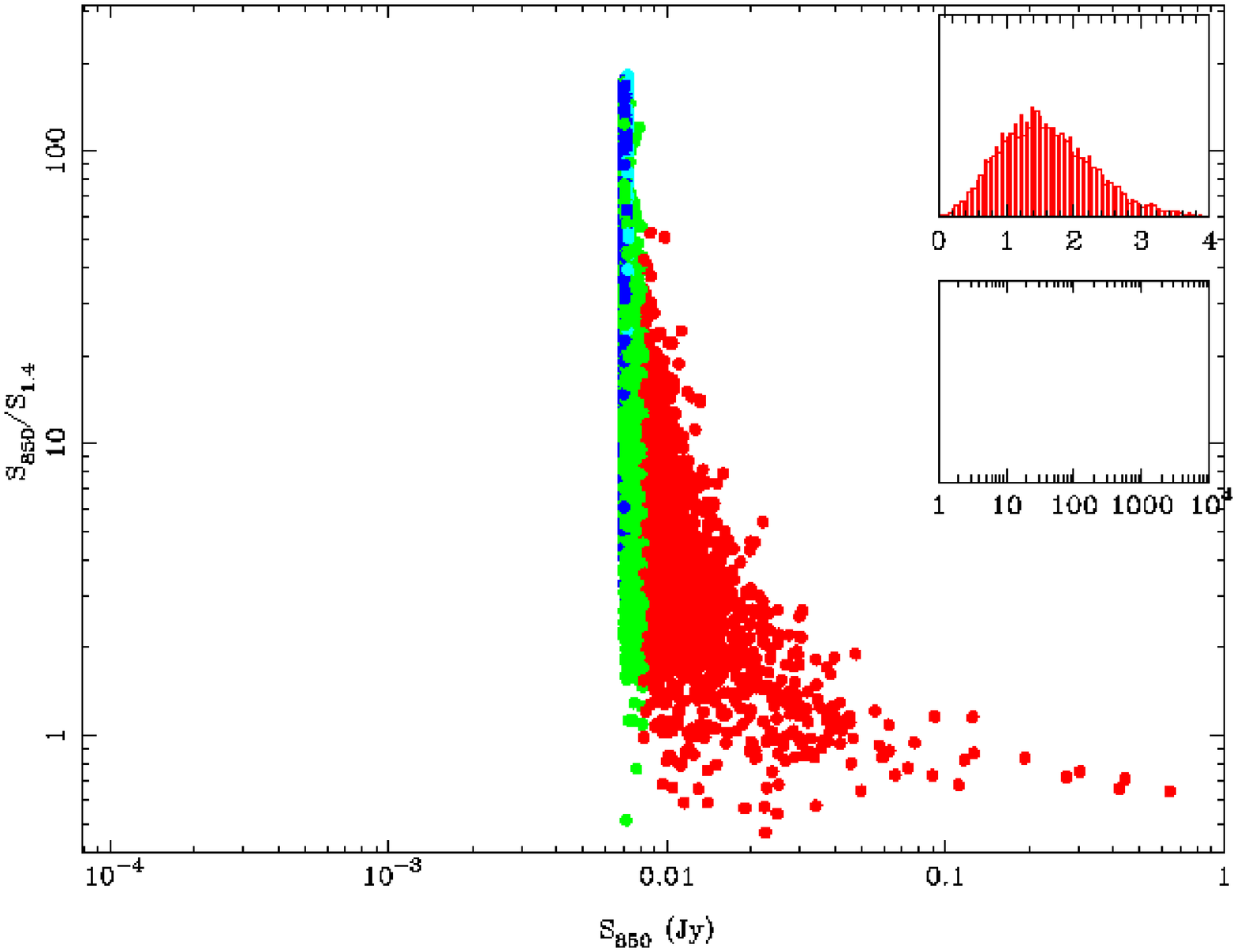,height=6.25cm,angle=-90}
\epsfig{file=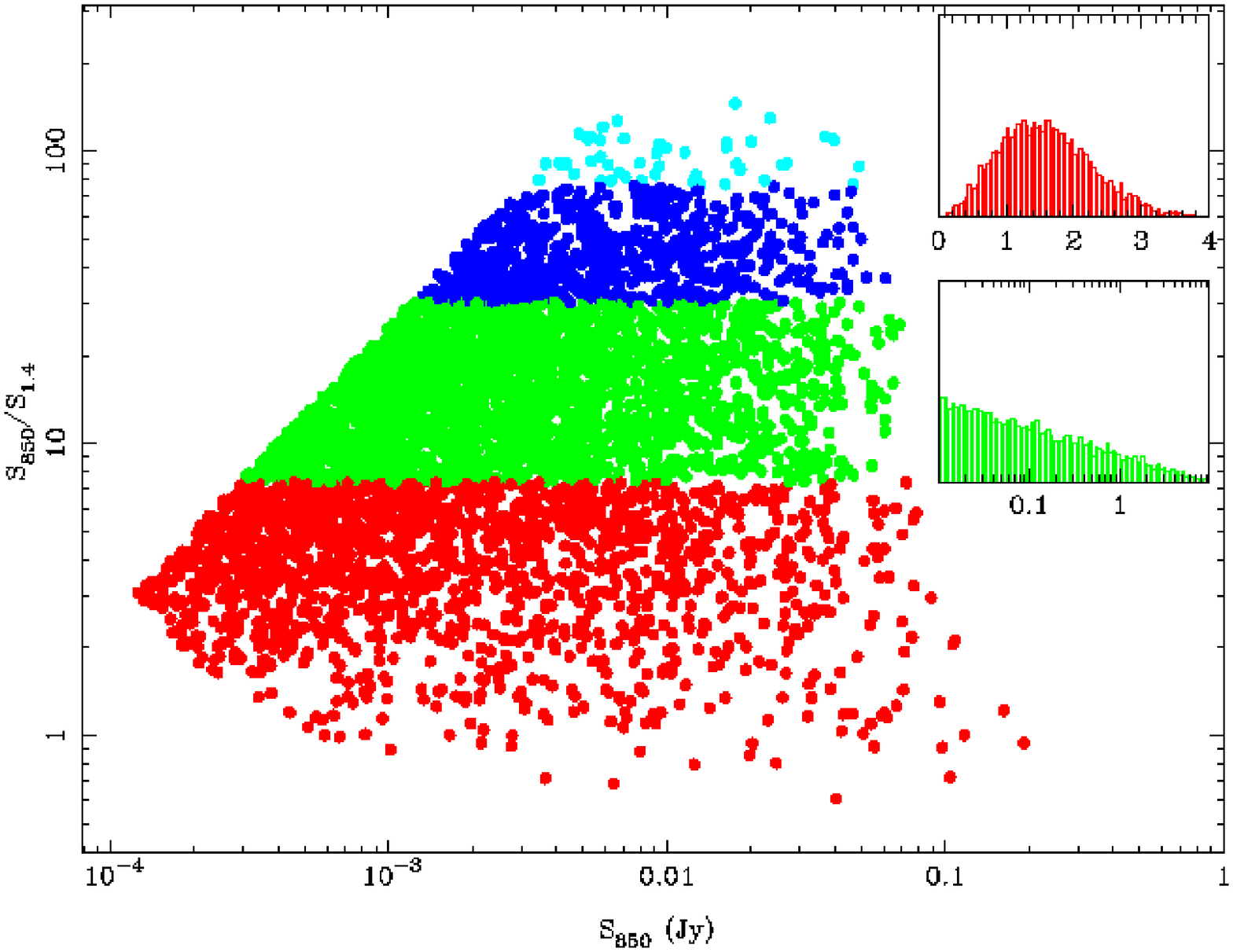,height=6.25cm,angle=-90}
\epsfig{file=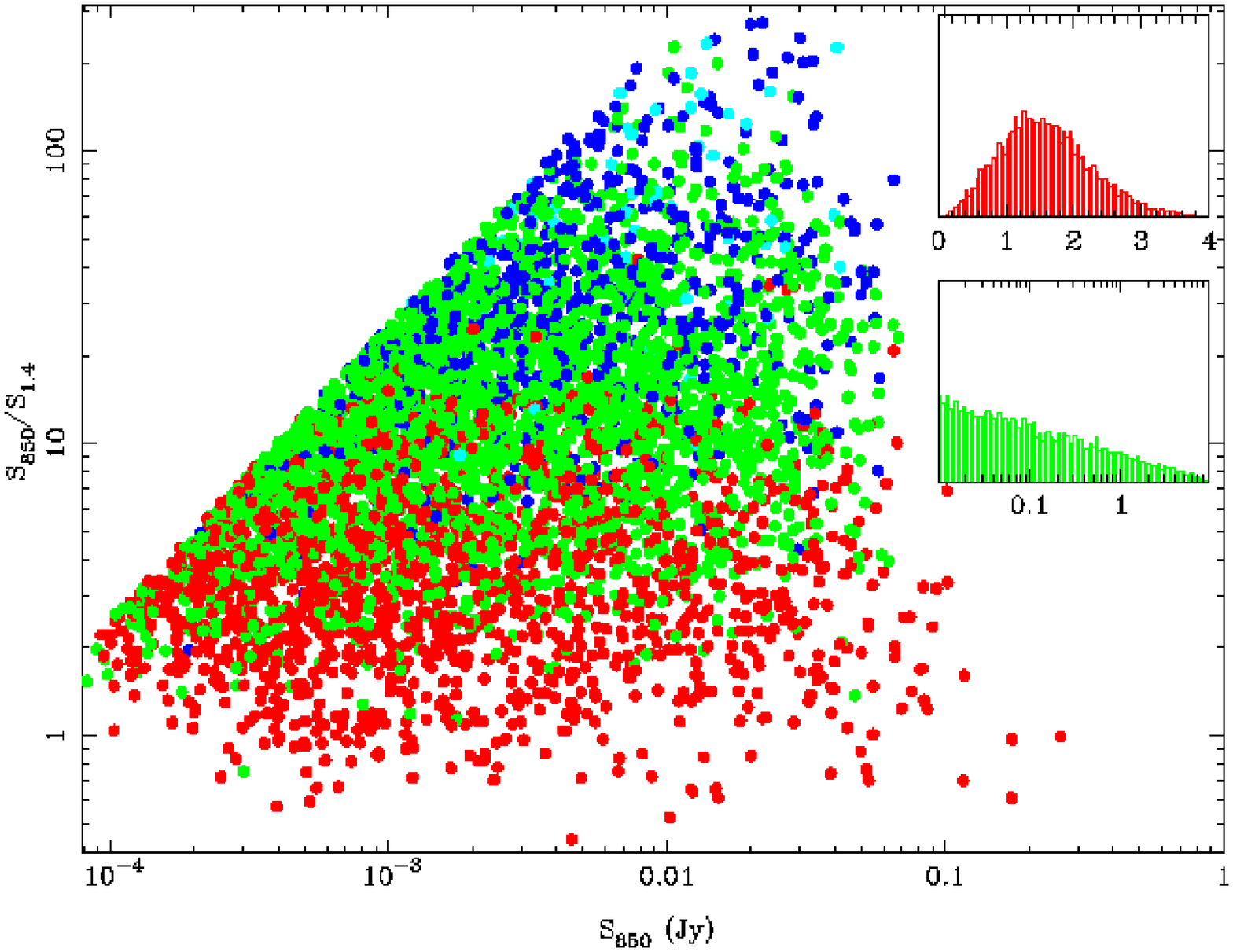,height=6.25cm,angle=-90}
\end{center}
\label{X}
\figcaption[chapman.fig19]{
Monte Carlo simulations of the 850\mum/1.4\,GHz CMD, showing the
null hypothesis for pure luminosity evolution.
{\bf upper left:} A population of ULIRG galaxies with about twice the
luminosity of Arp220 is plotted, following
a Gaussian redshift distribution (shown in the inset). 
The cyan, blue, green and red color bands correspond to $z\,{>}\,3$,
$2\,{<}\,z\,{<}\,3$, $1\,{<}\,z\,{<}\,2$, and $z\,{<}\,1$ respectively.
The corresponding track through the CMB shows the redshift spread of
galaxies of fixed luminosity and SED.  This becomes vertical at high $z$,
but redshifts are mapped one-to-one onto a specific value of
$S_{850}$/$S_{1.4}$. 
{\bf upper right:} Now we introduce a spread in the $S_{850}$/$S_{1.4}$ 
ratio, corresponding to the locally observed scatter as measured in 
Table~1.
{\bf lower left:} Imposing a luminosity distribution on our
ULIRG population, represented here simply as a power law (shown in the
second inset), is probably a reasonable
assumption for the bright sources considered here.
At this point, the radio survey limit imposes a diagonal
cut out of the upper left corner of the CMD.
{\bf lower right:} The effects of redshift distribution, sub-mm/radio scatter,
and luminosity function are combined to create a possible realization
for the CMD relation.  This scenario is clearly incompatible with our
observed CMD.
}
\end{minipage}
\end{figure*}

%
%
%
\begin{figure*}
\begin{minipage}{170mm}
\begin{center}
\epsfig{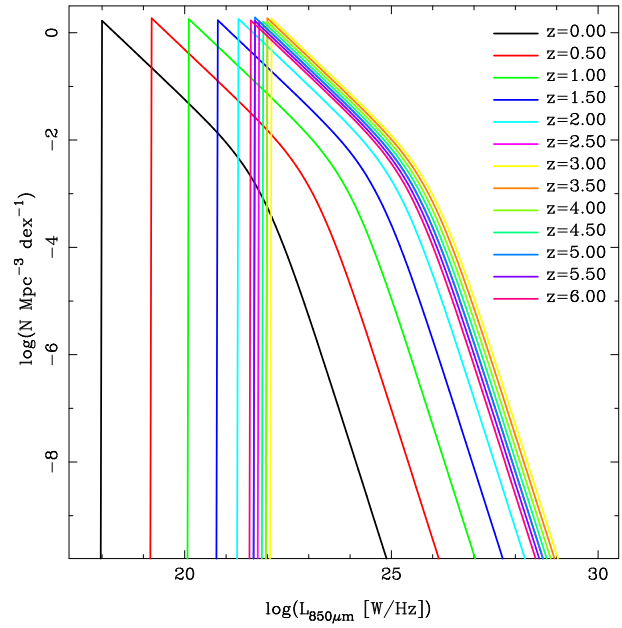}
\end{center}
\label{X}
\figcaption[chapman.fig18]{
The luminosity functions adopted in our Monte Carlo models are
presented, taking the form of the local {\sl IRAS\/} 60\mum\ function 
(Saunders et al.~1990) mapped to 
850\mum\ using the best fit dust temperature and emissivity from 
Dunne et al.~(2000a).  The evolution is pure luminosity, taking the
form $(1+z)^{4}$, consistent with the Blain et al.~(1999a) modeling 
results.  Beyond $z\,{=}\,3$ the evolution falls with a form $(1+z)^{-4}$.
The lines are color coded by redshift as noted in the figure.
Note that the truncation of the function at low luminosities is irrelevant
for the flux densities probed by SCUBA.}

\end{minipage}
\end{figure*}

%
%
%
\begin{figure*}
\begin{minipage}{170mm}
\begin{center}
\epsfig{file=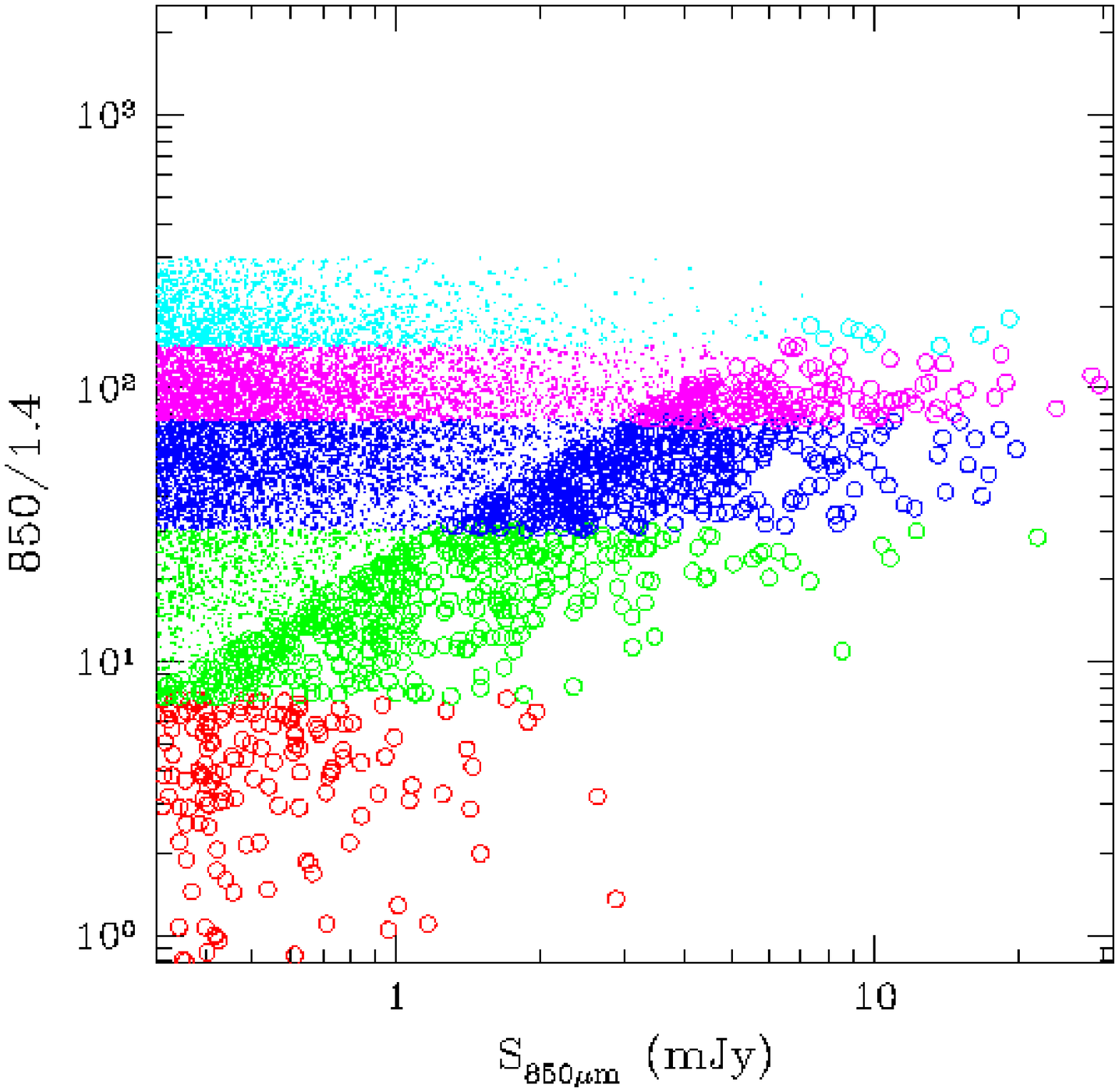,height=11.25cm,angle=0}
\epsfig{file=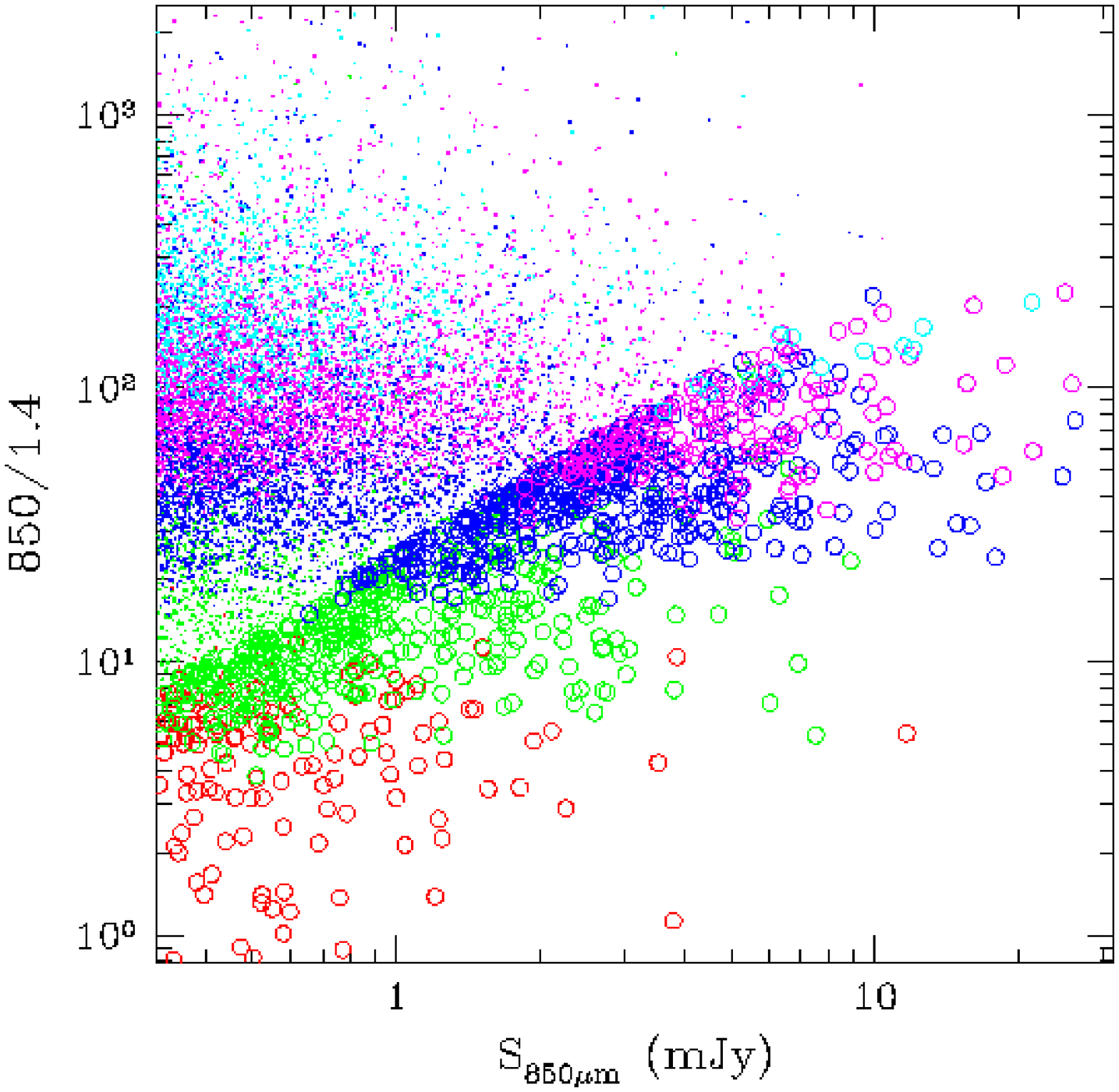,height=11.25cm,angle=0}
\end{center}
\label{X}
\figcaption[chapman.fig19]{
Monte Carlo models including luminosity evolution.
{\bf upper:} Simulation of the 850\mum/$1.4\,$GHz CMD, but
without the sub-mm/radio scatter for clarity.  For illustration of the 
total sample and the effect of radio pre-selection, we show all sources,
but those we can detect have larger symbols.
The distribution of detectable sources
is qualitatively similar to our observed CMD.
Redshifts are color-coded: red $z\,{<}\,1$; green $z\,{=}\,1$--2;
blue $z\,{=}\,2$--3; magenta $z\,{=}\,3$--4, cyan $z\,{>}\,4$.
{\bf lower:}
Scatter in the sub-mm/radio relation is now added.
The source density for $S_{850}\,{>}\,5\,$mJy matches that of
our observed CMD.
}
\end{minipage}
\end{figure*}

%
%
%
\begin{figure*}
\begin{minipage}{170mm}
\begin{center}
\epsfig{file=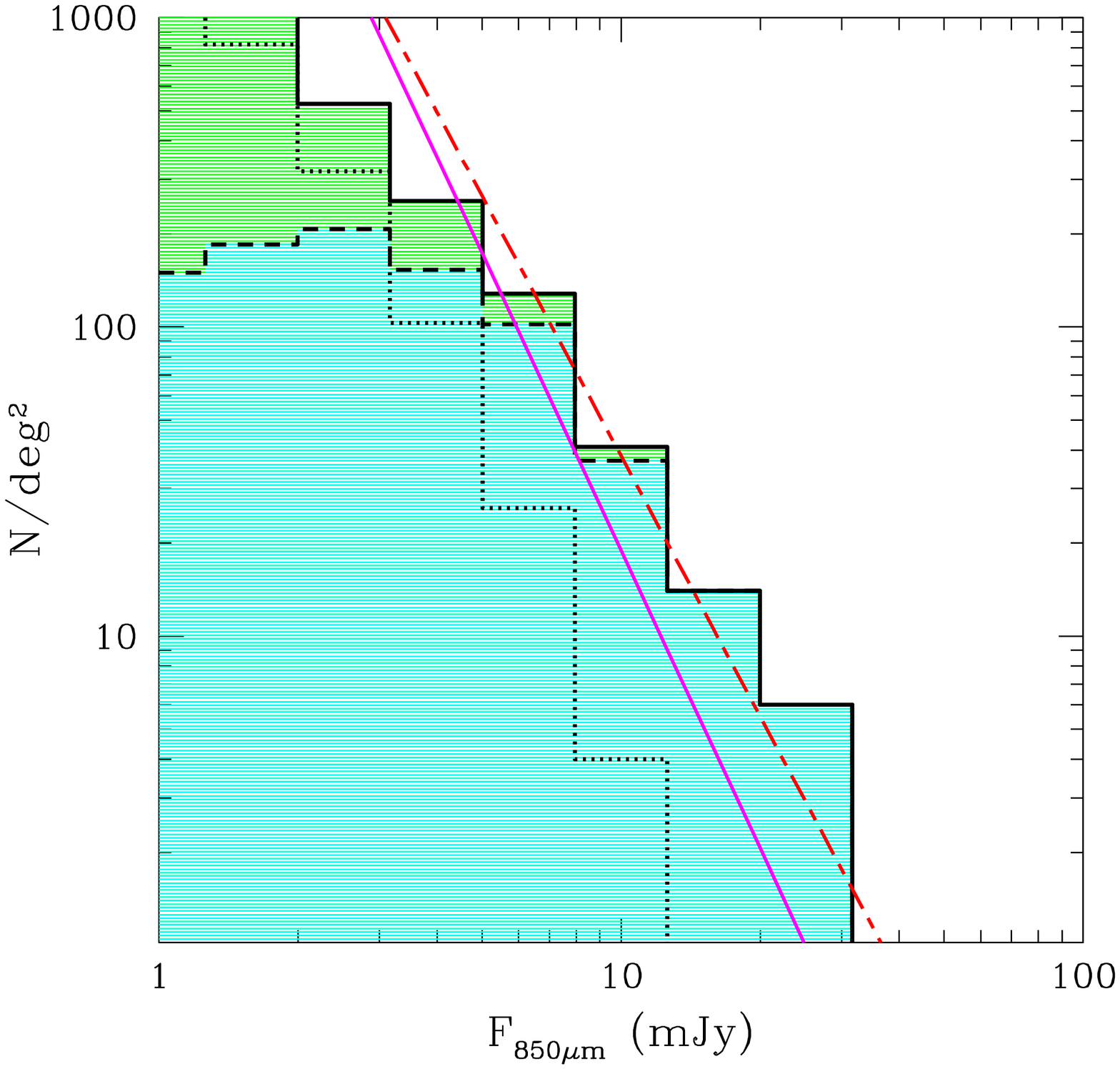,height=10.25cm,angle=0}
\epsfig{file=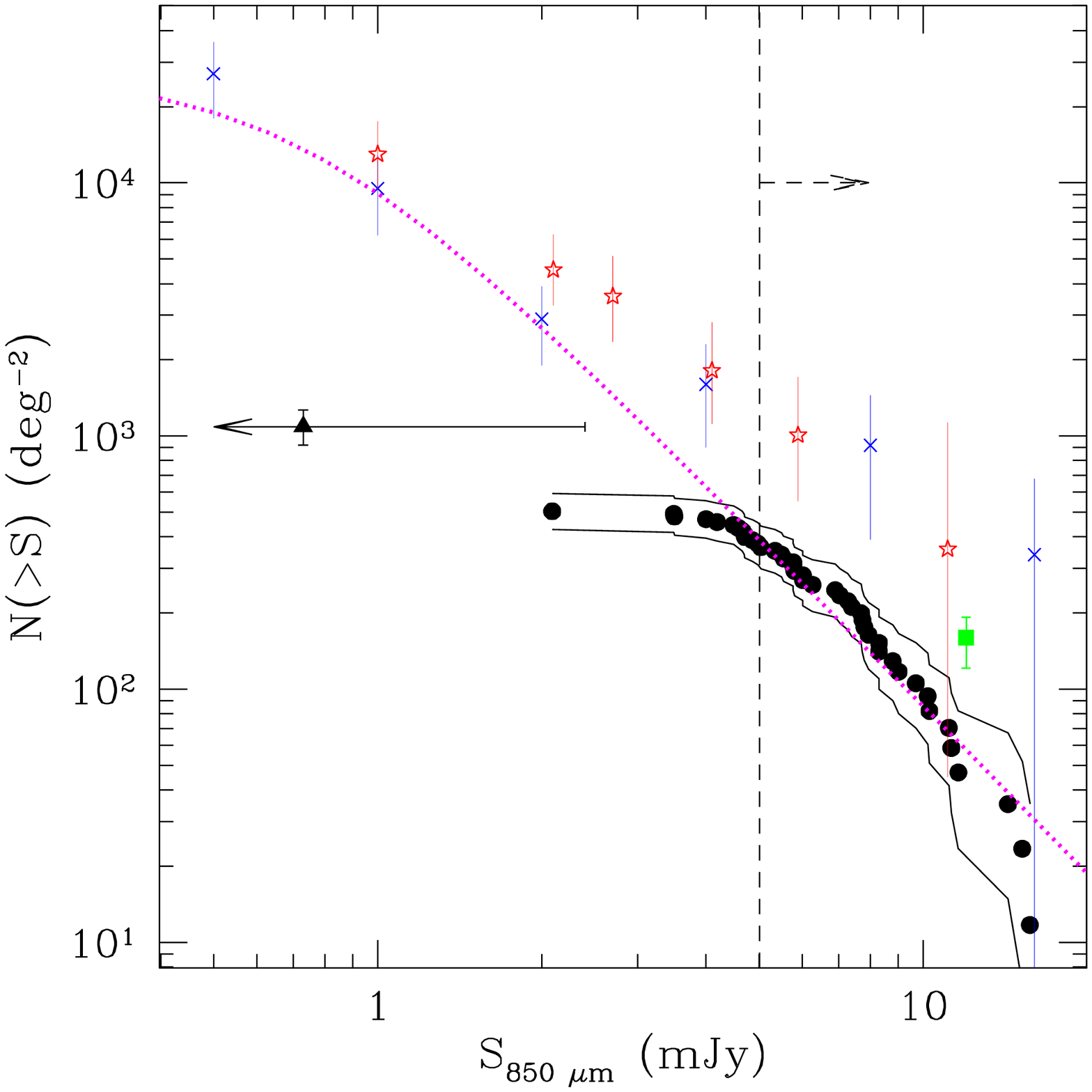,height=10.25cm,angle=0}
\end{center}
\label{X}
\figcaption[chapman.fig19]{
{\bf Upper panel:}
Differential source counts extracted from our Monte Carlo simulations. 
By binning the sources from our CMD as a function of $S_{850}$, we can
recover the total source counts, as well as the fraction detected in the
radio.  The solid red line represents the power law fit to the blank field
sub-mm source counts (Barger et al.~1999), while the dashed
magenta line represents a fit to the bright end of the
lensed counts of Blain et al.~(1999) and Chapman et al.~(2002c). 
The histograms represent the total modeled counts (solid), the radio
detected counts (dashed), and the radio undetected counts (dotted).
With our adopted scatter of 0.2 dex in the sub-mm/radio correlation, we recover
a large percentage of the sub-mm counts brighter than 10\,mJy, falling to
a 50\% recovery rate by ${\sim}\,5\,$mJy.
{\bf Lower panel:}
The integrated source counts estimated from our
OFRS sample, shown by filled symbols, with estimated error band.
Also plotted is the fit to blank
field sub-mm survey counts of Barger et al.~(1999) and Eales et al.~(2000),
shown by the dotted line, and the lensing amplified survey
counts of Blain  et al.~(1999) and Chapman et al.~(2002c), shown as crosses
and stars respectively.  A recent measurement in the extended HDF region
for sources brighter than $12\,$mJy is shown by the square.
The fraction of sources recovered is clearly high.
}
\end{minipage}
\end{figure*}

%
%
%
\begin{figure*}
\begin{minipage}{170mm}
\begin{center}
\epsfig{file=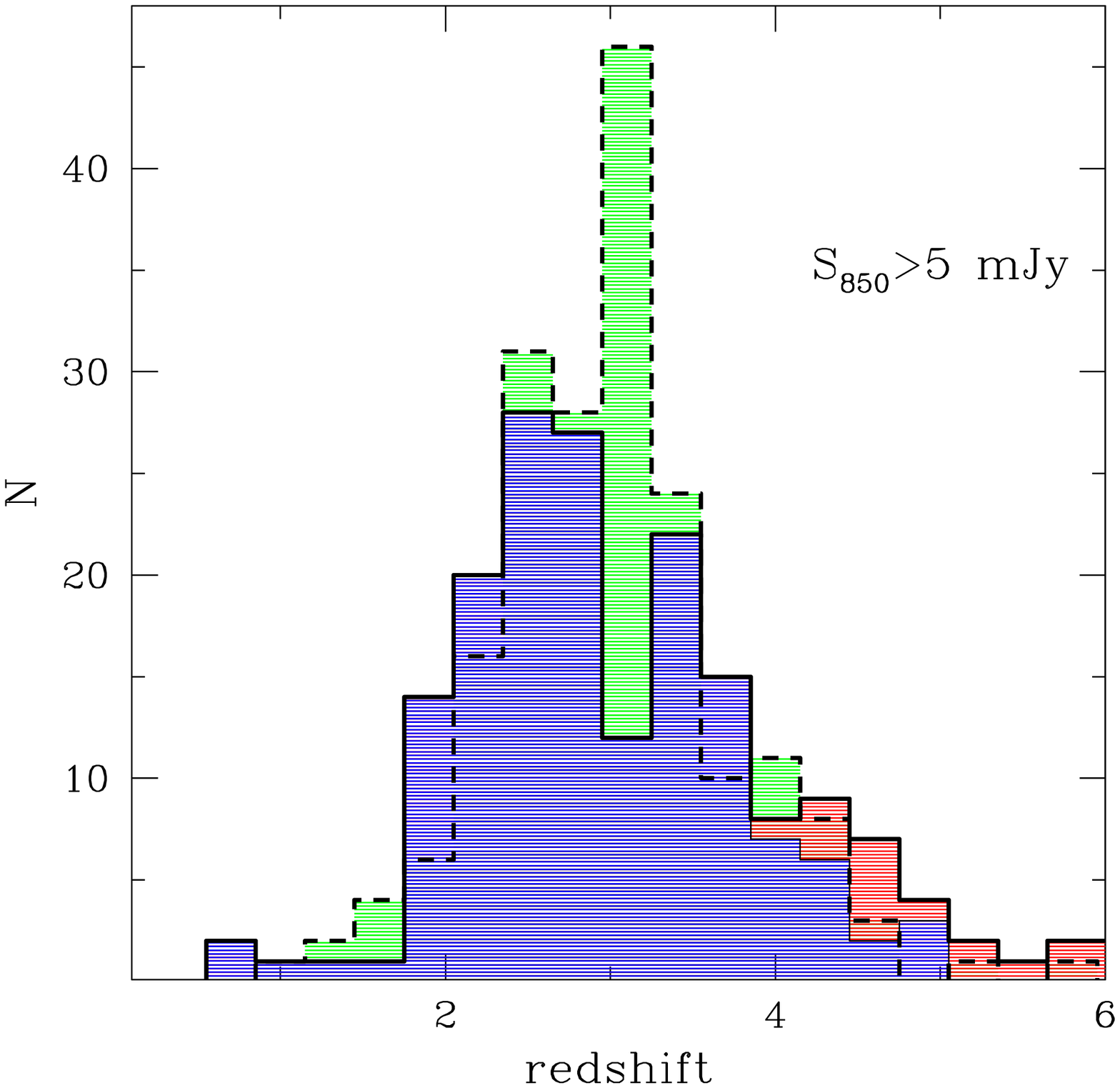,height=7.25cm,angle=0}
\epsfig{file=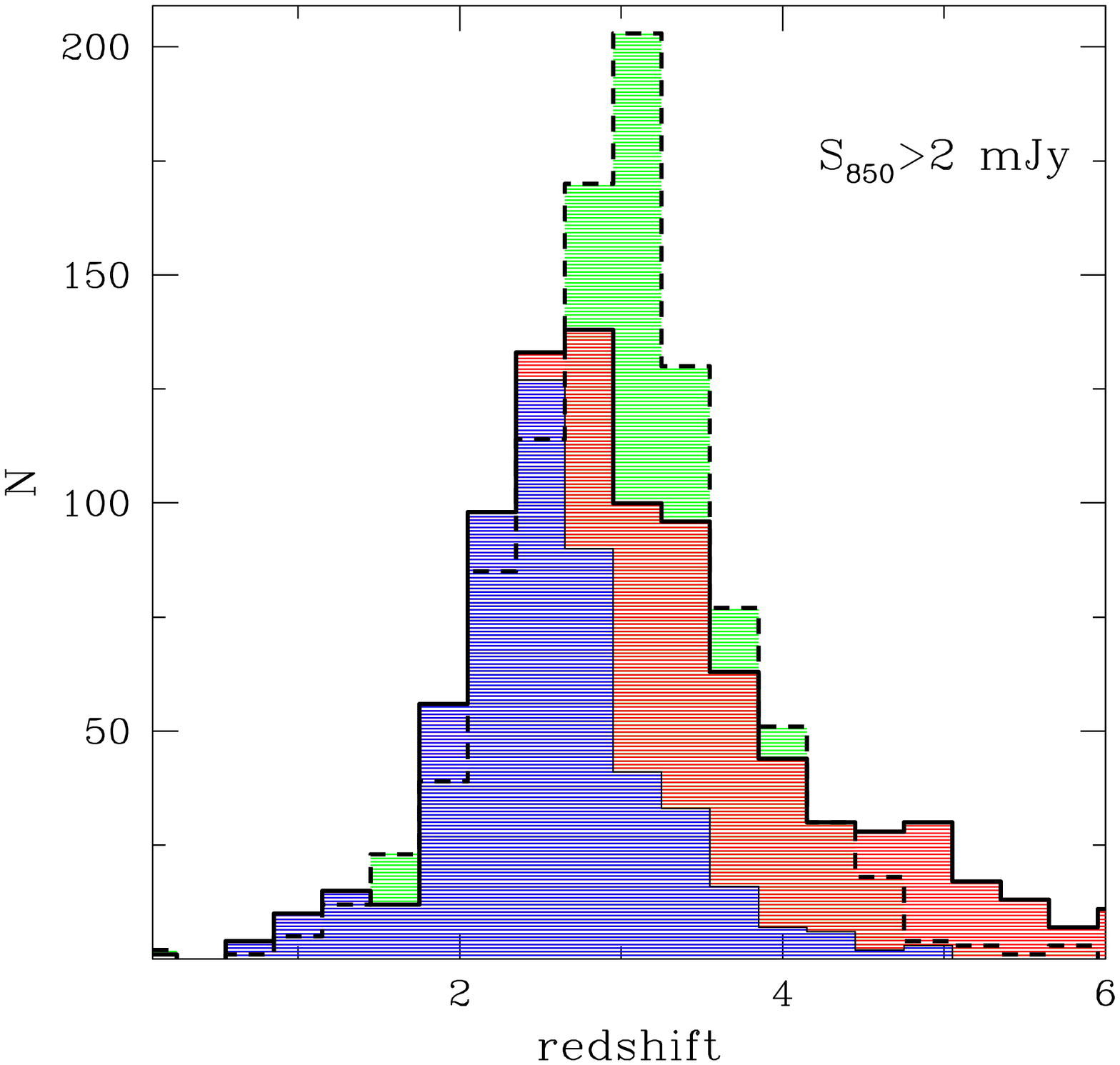,height=7.25cm,angle=0}
\epsfig{file=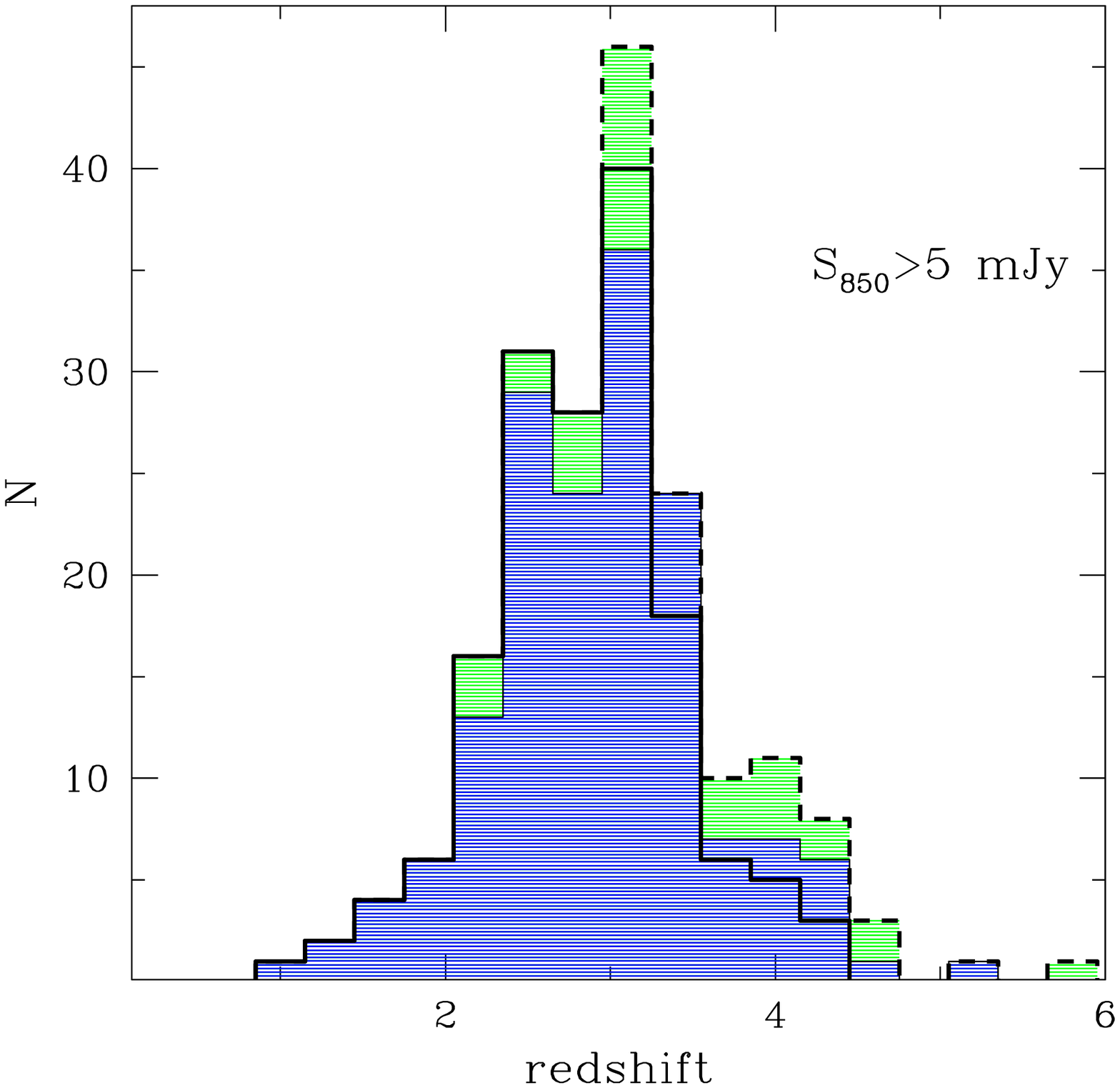,height=7.25cm,angle=0}
\epsfig{file=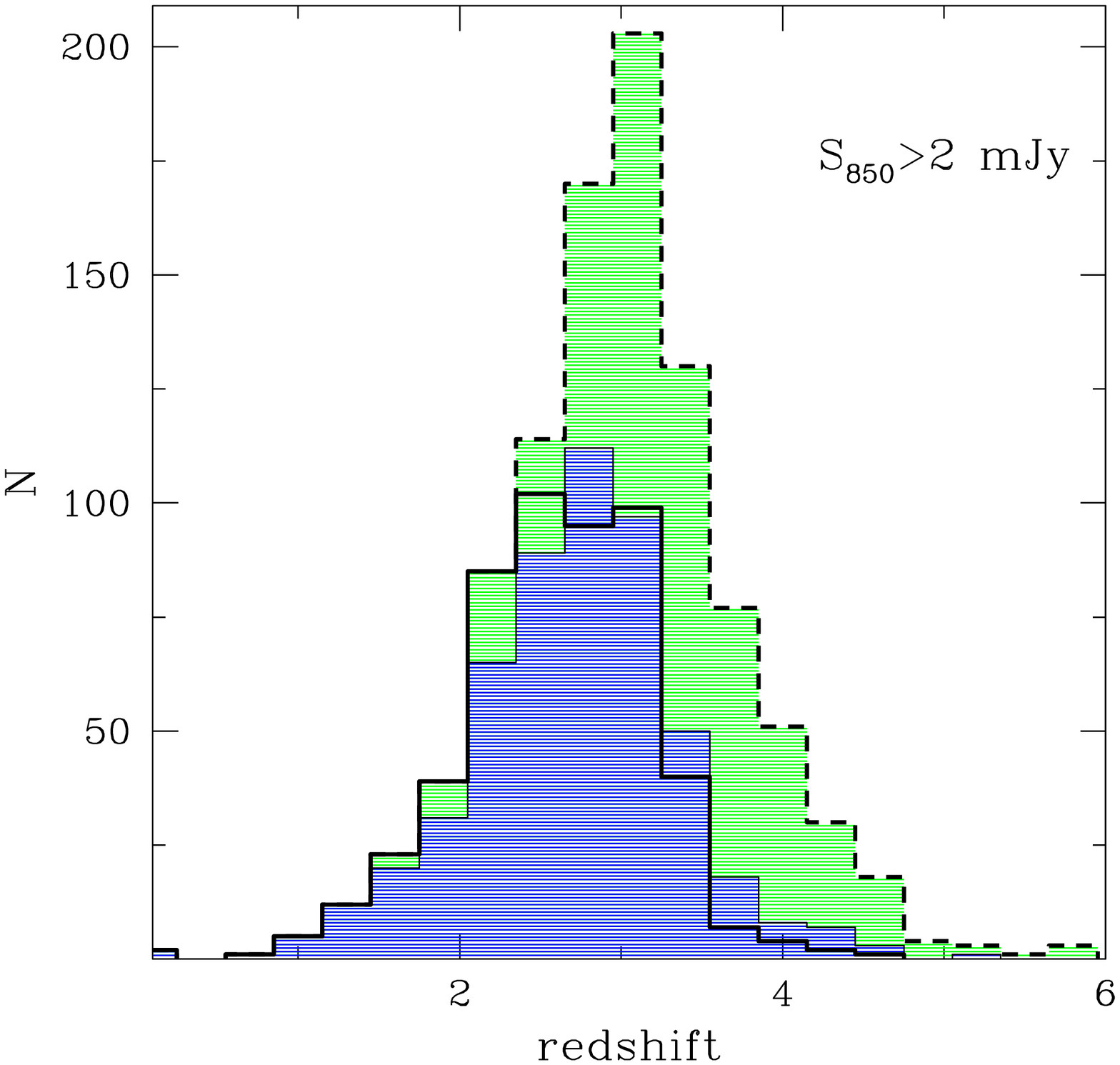,height=7.25cm,angle=0}
\epsfig{file=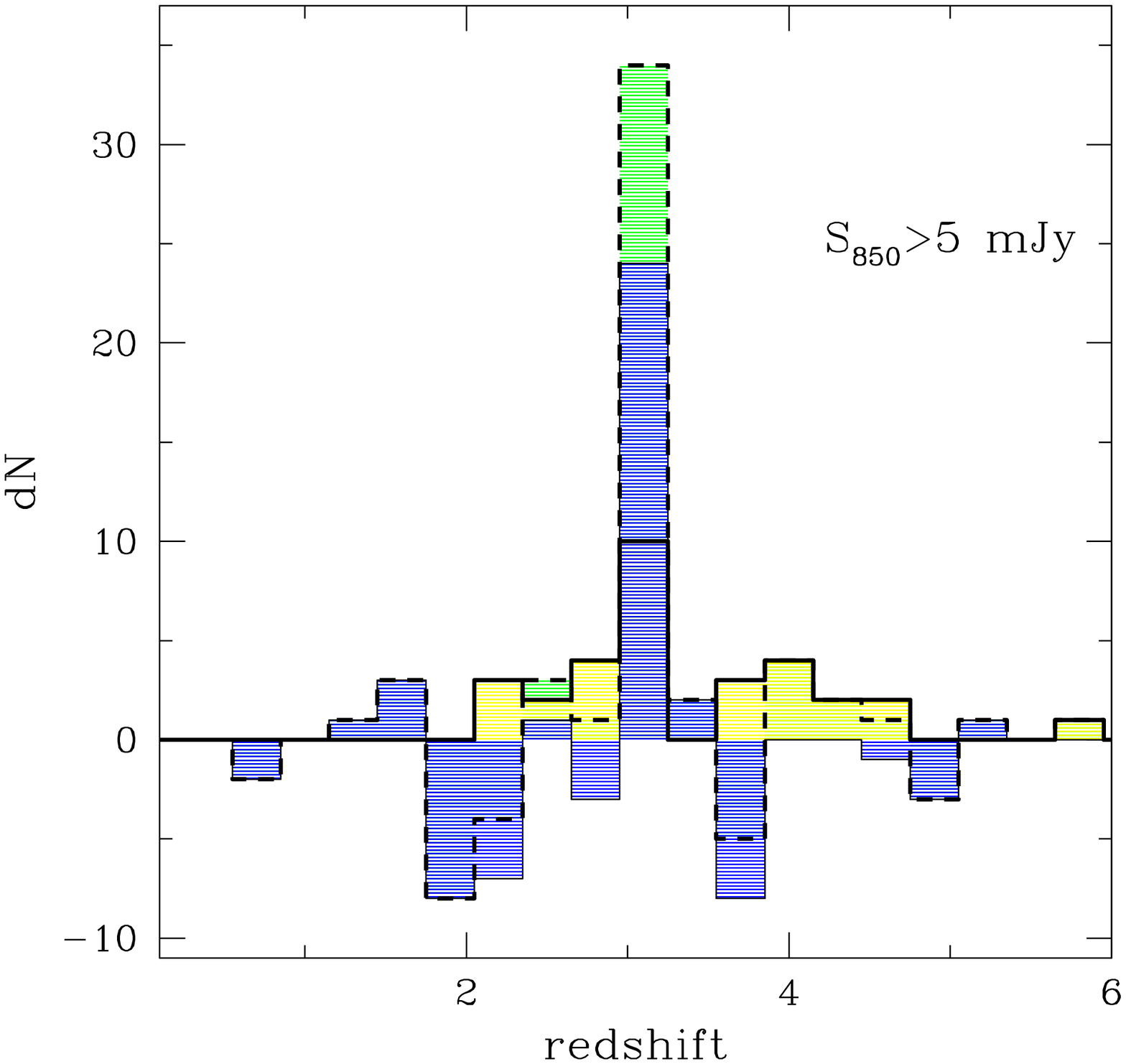,height=7.25cm,angle=0}
\epsfig{file=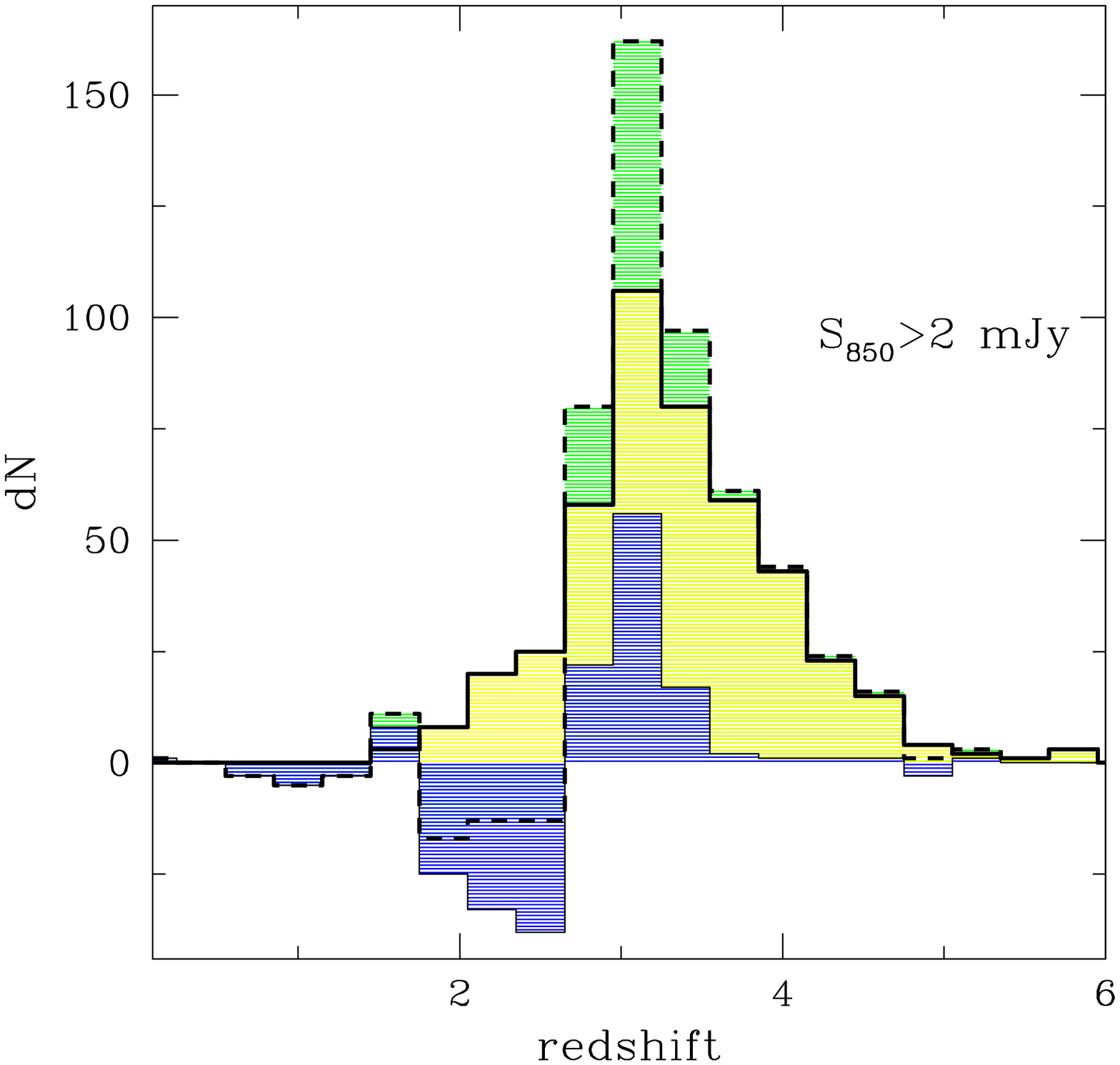,height=7.25cm,angle=0}
\end{center}
\label{X}
\figcaption[chapman.fig19]{
{\bf Top row:}
The modeled redshift distributions extracted from our Monte Carlo 
simulations.  Green histograms (dashed lines) represent the input redshift
distribution of the sources, while the distribution as measured
from the 850\mum/$1.4\,$GHz ratio is plotted as the blue/red histograms
(solid lines).
The blue shaded inset in the latter case represents the fraction
of sources recovered from the radio pre-selection.
The distribution measured from the 850\mum/1.4\,GHz (including 0.2 dex 
scatter) is flattened and also broader than the true distribution. 
There is an excess of sources 
it lower redshift, and also an excess of high-$z$ sources,
arising simply from the intrinsic scatter in the 850\mum/$1.4\,$GHz relation,
which broadens the peak in our adopted evolution function.
{\bf Middle row:} 
The actual
redshift distribution recovered by the radio selection (blue histogram)
compared to the recovery in the absence of scatter in the far-IR/radio
relation.  In our $5\,$mJy flux cut, we recover an excess of sources for all
$z\,{>}\,3$, but recover a deficit of sources from $2\,{<}\,z\,{<}\,3$.
Our radio selected survey is therefore less complete at
$2\,{<}\,z\,{<}\,3$ than we would have naively expected.
{\bf Bottom row:} The residuals as an estimate of the errors
involved in both the intrinsic radio selection and the 850\mum/$1.4\,$GHz
estimator. 
The dashed line (green) shows the residuals of the radio detected sources with
a 850\mum/$1.4\,$GHz estimated redshift distribution, compared with
the input distribution.
The solid line (yellow) shows the true redshifts of radio selected sources
relative to the total distribution.  The light solid line (blue)
shows the difference in the 850\mum/$1.4\,$GHz estimated redshift
from the true redshift distribution for the radio selected sources alone.  
}
\end{minipage}
\end{figure*}

%
%
%
\begin{figure*}
\begin{minipage}{170mm}
\begin{center}
\epsfig{file=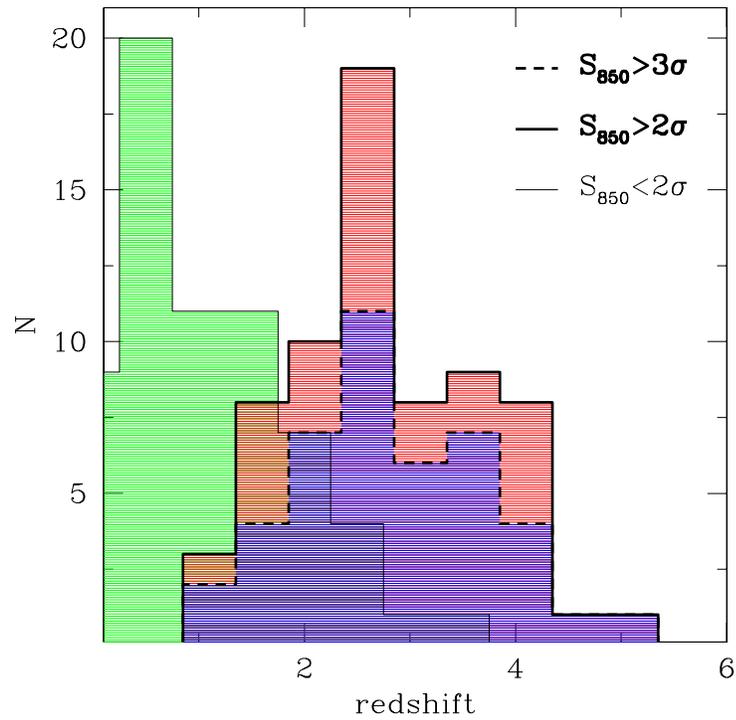,height=10.25cm,angle=0}
\end{center}
\label{X}
\figcaption[chapman.fig19]{The estimated redshift distribution for 
SCUBA detected and undetected radio sources in our sample.
We plot the histogram with a dashed (blue) line for radio sources detected
in the sub-mm above a $3\sigma$ limit.  We also plot the distribution
for all sources with $S_{850}\,{>}\,2\sigma$ (solid line, red histogram).
Sources with ${<}\,2\sigma$ are shown with a light solid line (green),
suggesting a tail of faint sub-mm sources extending to 
redshifts comparable to those spanned by the sub-mm detected sample.
This 850\mum/$1.4\,$GHz constructed redshift distribution is matched
by our Monte Carlo simulations (compare Fig.~8). 
}
\end{minipage}
\end{figure*}

%
%
%
\begin{figure*}
\begin{minipage}{170mm}
\begin{center}
\epsfig{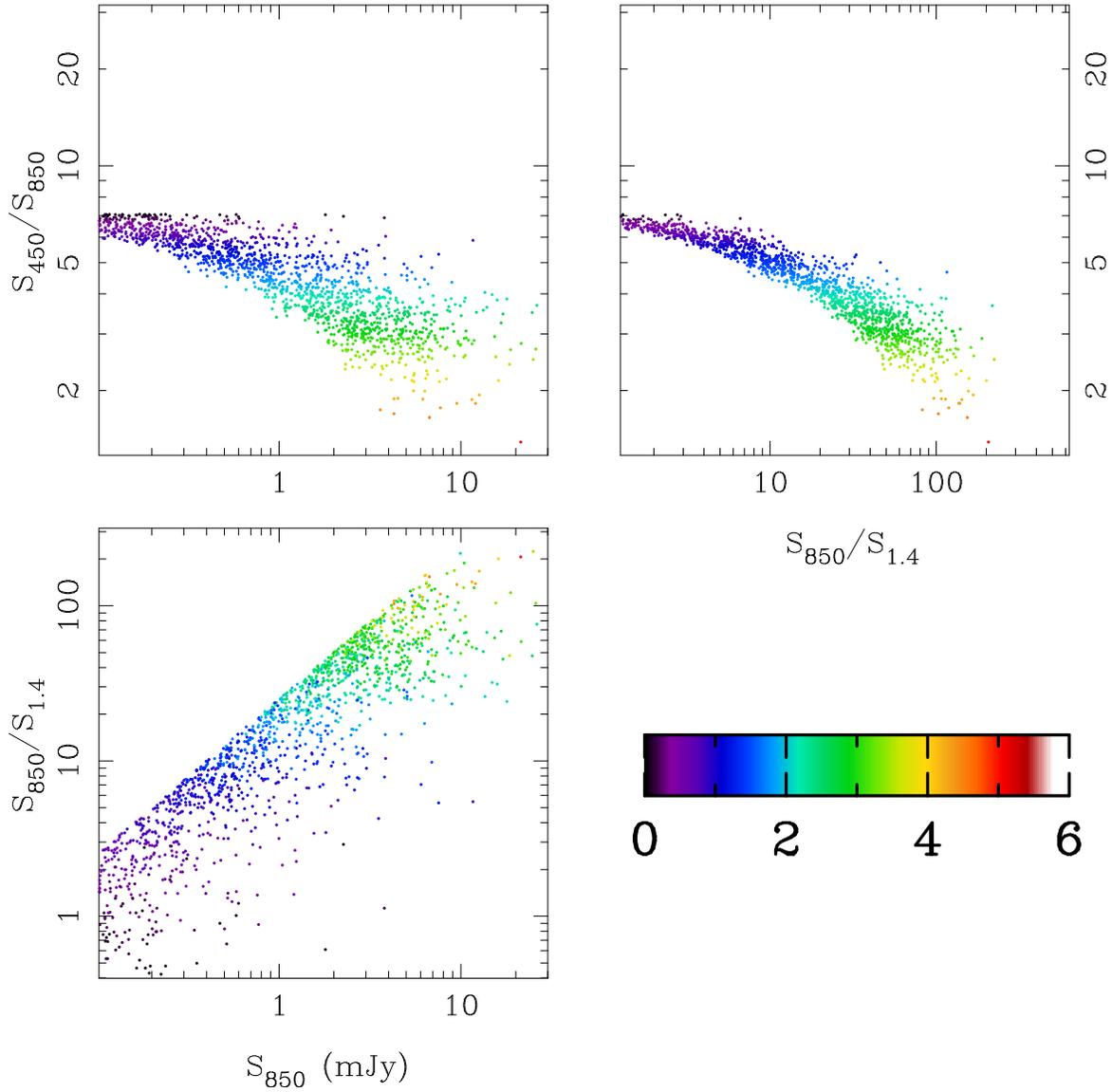}
\end{center}
\label{X}
\figcaption[chapman.fig10]{
450\mum\ band estimates can be extracted directly from our models,
which effectively use a single dust temperature of $50\,$K 
with an emissivity of $\beta=1.5$.  The model sources are color
coded by redshift as indicated on the bar at lower right.  The upper
left plot is the $S_{450}/S_{850}$ versus $S_{850}$ CMD.  The main
variable affecting the 450\mum/850\mum\ ratio is the redshift,
higher redshift sources having lower ratios.  The upper right plot shows
the $S_{450}/S_{850}$ versus $S_{850}/S_{1.4}$ color-color diagram, which
appears similar to the CMD, although somewhat tighter as a result of the
relation between $S_{850}/S_{1.4}$ and $S_{850}$, which we plot in
the lower left panel.
}
\end{minipage}
\end{figure*}

%
%
%
\begin{figure*}
\begin{minipage}{170mm}
\begin{center}
\epsfig{file=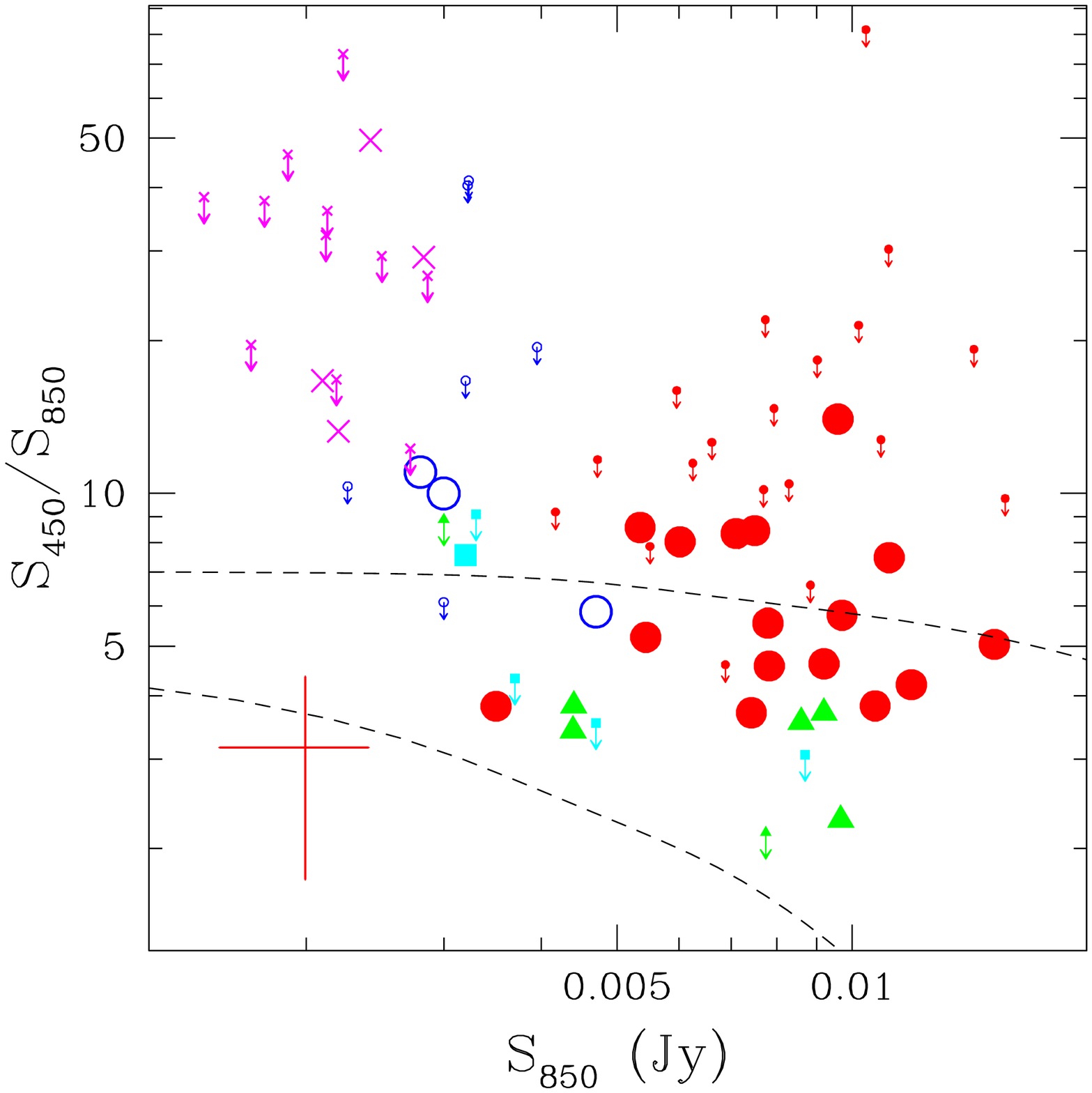,height=10.25cm,angle=0}
\epsfig{file=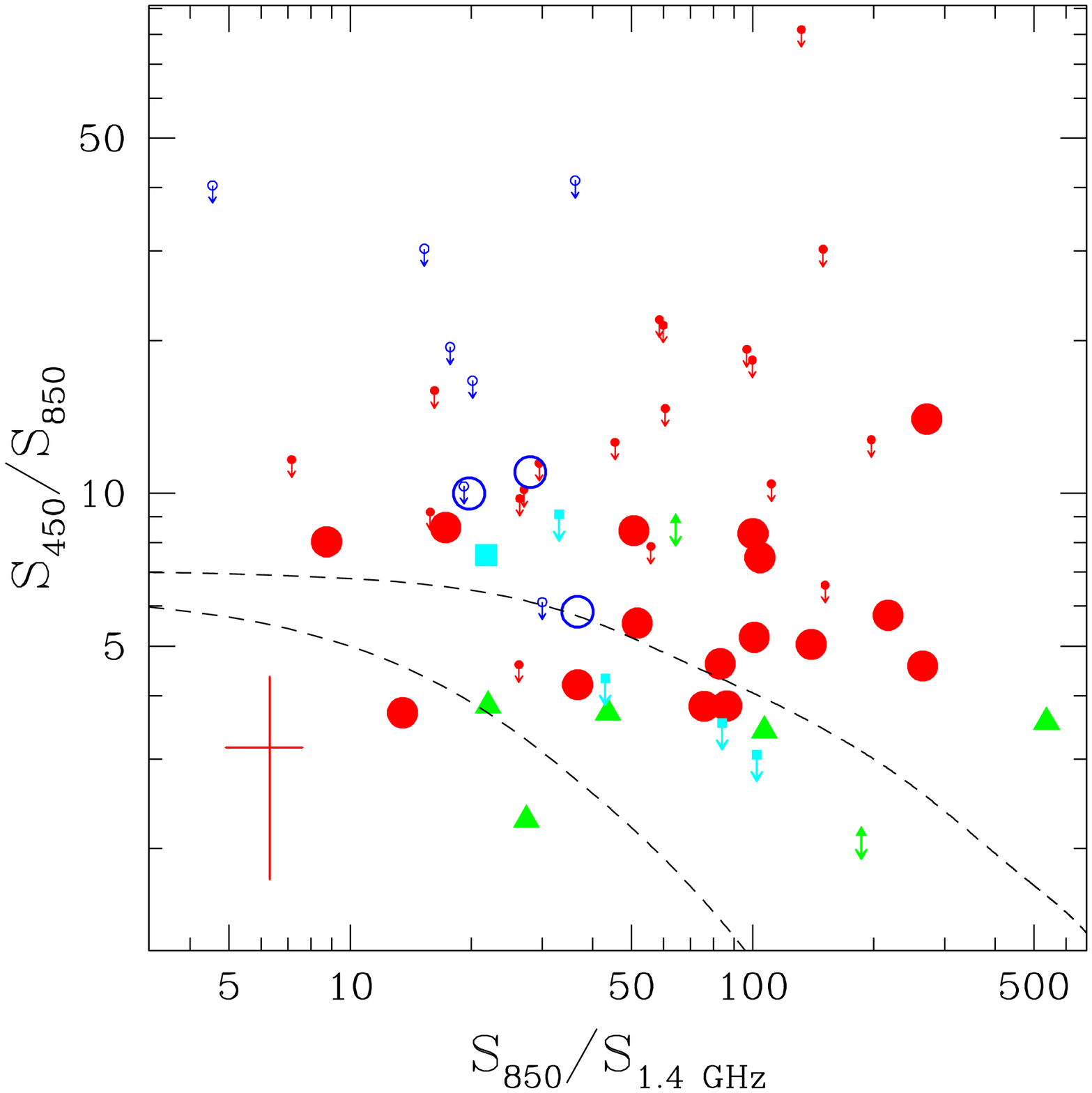,height=10.25cm,angle=0}
\end{center}
\label{X}
\figcaption[chapman.fig11]{
19 sources 
450\mum\ flux density detected at $>$2$\sigma$.
A representative error-bar
for our detected sample is displayed in the lower left corner.
The upper panel shows the $S_{450}$/$S_{850}$ versus $S_{850}$ CMD, with all
sources undetected at 450\mum\ presented as 2$\sigma$ upper limits. 
Dashed lines indicate the envelope of our models.
The lower panel presents the $S_{450}$/$S_{850}$ versus $S_{850}$/$S_{1.4}$
color-color diagram.
The results from our single (50\,K) dust temperature model are overlaid as an
envelope defining the region of detected sources from Fig.~10. 
The measured sources are consistent within $1\sigma$ of the 
envelope, but on average have a higher 450\mum/850\mum\ value.
The sources from the Smail et al.~(2001) catalog (triangles)
and the source from Eales et al.~(2000) (square)
lie near the lowest values of 450\mum/850\mum\ that we find
from the OFRS sample (red circles).
Given the relation between $S_{850}$/$S_{1.4}$ and
$S_{850}$, the color-color plot is similar to the CMD. The 
source positions are rearranged in the color-color plot, 
suggesting that the 850\mum\ 
luminosities are somewhat independent of redshift.
}
\end{minipage}
\end{figure*}

\end{document}